\documentclass[%
 reprint,
 amsmath,amssymb,
 aps,
 floatfix,
 showpacs
]{revtex4-2}

\usepackage{graphicx}
\usepackage{dcolumn}
\usepackage{bm}
\usepackage[dvipsnames]{xcolor}
\usepackage{graphics}
\usepackage{epsfig}
\usepackage{color}

\begin{document}

\preprint{APS/123-QED}

\title{Chase-Escape Percolation on the 2D Square Lattice}

\author{Aanjaneya Kumar}
\email{kumar.aanjaneya@students.iiserpune.ac.in}
\affiliation{Department of Physics, Indian Institute of Science Education and Research, Dr. Homi Bhabha Road, Pune 411008, India}

\author{Peter Grassberger}
\email{p.grassberger@fz-juelich.de}
\affiliation{JSC, Forschungszentrum Jülich, D-52425 Jülich, Germany}

\author{Deepak Dhar}
\email{deepak@iiserpune.ac.in}
\affiliation{Department of Physics, Indian Institute of Science Education and Research, Dr. Homi Bhabha Road, Pune 411008, India }

\date{\today}

\begin{abstract}
Chase-escape percolation is a variation of the standard epidemic spread models. In this model, each site can be in one of three states: unoccupied, occupied by a single prey, or occupied by a single predator. Prey particles spread to neighboring empty sites at rate $p$, and predator particles spread only to neighboring sites occupied by prey particles at rate $1$, killing the prey particle that existed at that site. It was found that the prey can survive with non-zero probability, if $p>p_c$ with $p_c<1$. Using Monte Carlo simulations on the square lattice, we estimate the value of $p_c = 0.49451 \pm 0.00001$, and the critical exponents are consistent with the undirected percolation universality class. We define a discrete-time parallel-update version of the model, which brings out the relation between chase-escape and undirected bond percolation. For all $p < p_c$ in $D$-dimensions, the number of predators in the absorbing configuration has a stretched-exponential distribution in contrast to the exponential distribution in the standard percolation theory. We also study the problem starting from the line initial condition with predator particles on all lattice points of the line $y=0$ and prey particles on the line $y=1$. In this case, for $p_c<p < 1$, the center of mass of the fluctuating prey and predator fronts travel at the same speed. This speed is strictly smaller than the speed of an Eden front with the same value of $p$, but with no predators. At $p=1$, the fronts undergo a depinning transition. The fluctuations of the front follow Kardar-Parisi-Zhang scaling both above and below this depinning transition.
\end{abstract}

\maketitle


\section{Introduction}

Growth and spreading processes are ubiquitous, e.g. growing crystals from supersaturated solutions \citep{sol1, sol2}, 
dielectric breakdown \citep{db1, db2}, bacterial colonies \citep{bact}, growth of tumors \citep{tumor2, tumor} and spreading 
of rumours in social networks \citep{cer, rumor, retract}. Spreading processes and stochastic growth models have attracted 
a lot of attention in statistical physics, owing to their wide applicability and their beauty \citep{grow1,hamfpp,eden,50fpp}. 
Simple mathematical models of these have provided us with much insight about their universal critical behavior, and more 
generally into nonequilibrium phenomena  \citep{kpz, kpz2,araujo,hh1}. One of the earliest of such models was the Eden model 
of tumour growth \citep{eden}.  Many questions about this model have been studied, such as the velocity of its growth in 
different directions, the  asymptotic shape  \citep{dhar-eden, wolf, bertrand-pertinand,eden2}, and fluctuations of the interface \citep{50fpp}. 
Several variants have been studied, starting from the model of skin cancer by Williams and Bjerknes \citep{skin} to the SIR 
(Susceptible-Infected-Removed) and SIS (Susceptible-Infected-Susceptible) models of epidemics \citep{sirs}. The importance 
of such modelling has been underscored during the current COVID-19 pandemic, where specialized SIR-type models have been 
extensively used to better understand the course of the pandemic, and study the effects of different intervention strategies 
\citep{adharind}.

An important issue in ecology and conservation studies is understanding how much predation would result in prey populations 
becoming extinct. Similarly, on social networks, one would like to know how efficient the rumor-scotching process has to be, 
in order to stop the spread of misinformation and rumors \citep{cer}. A simple model of this, called
\emph{Chase-Escape} percolation was introduced recently, where it was shown that indeed a slowly growing prey population 
can coexist with relatively fast growing predators. This is the subject of this paper. This model was first studied  in 
\citep{ce0} as a prey-predator model on trees where the prey tries to escape predators that are chasing it. Later, it was 
extended to lattices and graphs with cycles where the model displays very rich features \citep{ce2, ce3}. It was shown in 
Ref. \citep{ce2} that Chase-Escape  (CE) percolation on the $2D$ square lattice undergoes a survival-extinction 
phase transition for the prey, and interestingly, the critical point for this phase transition was shown to 
be very close to  $1/2$ \citep{ce2}, the 
well-known  critical threshold of bond percolation on the square lattice \citep{kesten}.  In fact, in \citep{ce3}, it was suggested that  the critical value for CE percolation on the square lattice may  also be exactly $1/2$. 

While there is no real argument in favor of the critical value for CE percolation on the square lattice being exactly 
$1/2$, it suggests a subtle connection to the  well-studied (Bernoulli)  bond percolation process (hereafter referred to  as the `standard'  percolation process).  One of the aims of our study was to determine 
the critical threshold in this model more precisely numerically by simulations. We find  $p_c =0.49451 \pm 0.00001$, which  rules out an exact correspondence between these two problems.    However, we will describe a discrete-time two-parameter generalization  of the CE process, which brings out  the connection to the standard percolation process. We further show that the entire subcritical phase has a stretched exponential decay of the cluster size distribution, as compared to 
the exponential distribution obtained in the case of standard percolation.  Nevertheless 
we also show that the critical behavior is completely consistent with being in the standard percolation universality class. 
In particular, the correlation length exponent $\nu$ obtained from a finite size scaling analysis and dynamical 
exponents obtained from cluster growth exactly at $p_c$ give values in complete agreement with the known $2D$ 
percolation values, and, using two different methods, we show that rotational invariance is recovered at the 
critical point $p_c$.

 In the range $p_c < p< 1$, we find that the velocity of the prey front is less than what it would be in the absence 
of  predators. This is different from the behavior found in some continuum predator-prey evolution models, where predators are 
unable to affect the front speed \citep{predation}.  We explain the difference in terms of the discrete nature of the front, which 
is not well-captured in the continuum description. Finally, we study the depinning transition between the predator and prey fronts 
at $p=1$, and estimate the roughness and dynamical exponents to be $\alpha = 1/2, z = 1.5$  in the entire range $p>p_c$, 
which is consistent with the KPZ universality class. 

The plan of the paper is as follows: in the next section, we  define the model precisely,  and review earlier results. In Section~III, we discuss our Monte Carlo simulation procedure and  the analysis -- based on `seeds' where we start with entire
lines of prey and predators -- which allows us to estimate the critical value 
$p_c$. Finite size scaling analysis for different lattice sizes is used to determine the critical exponent $\nu$ corresponding to 
the divergence of correlation length away from $p_c$. Simulations with point seeds are discussed in Sec.~IV, which give an even 
more precise estimate of $p_c$ and estimates for critical exponents at $p=p_c$. In Sec.~V, we discuss  a discrete-time version of the model, and determine the qualitative phase diagram. We find that one end-point of the line of CE transitions is exactly the standard bond percolation critical point on the square lattice. If the critical behavior is the same all along this line, this implies that it is in the percolation universality class. Some other phase boundaries corresponding to pinning-depinning transitions in this generalized model, that are in the directed percolation universality class are also determined exactly. 
In Section~VI, we discuss, for the general $D$-dimensional CE 
percolation, the subcritical regime  with  $ p < p_c$,  where all the prey particles eventually die out and the system enters one 
of its many absorbing states. We argue that the probability that of the number of predators in the absorbing configuration is 
greater than $s$  is bounded from below by $ \exp( -K p^{-1} s^{1/d})$, where $K$ is some $p$-independent constant.  This is 
in contrast to the usual percolation problem, where such distributions decay exponentially with $s$, for large $s$. In Section~VII, 
we simulate clusters starting with point seeds, which allows us to estimate critical exponents at $p=p_c$. Also, from these and
by means of growth starting from tilted line seeds, we study if the critical threshold is direction dependent, as is known to 
happen in the well-studied cases of directed percolation and Eden growth. We present evidence that it is not, again in agreement
with standard undirected percolation. Section~VIII is devoted to the second phase transition in this model, which is like a 
pinning-depinning transition of the predator and prey fronts, and occurs at $p=1$, but has not been discussed much earlier. 
For $p<1$, the prey and predator fronts move together at the same velocity. Our simulations show that these  pinned-together 
fronts travel slower than the Eden front moving at the same rate $p$,  in contrast with the 
results of Owen and Lewis for some continuum descriptions of prey-predator models \citep{predation}. We identify the mechanism for this slowing down of the prey front, and verify it using numerical simulations. Finally, we use finite-size scaling to show that both above and below the depinning transition, moving fronts are in the KPZ universality 
class. In Section~IX, we summarize our results and discuss some directions for further work. 

\begin{figure}
         \centering
         \includegraphics[scale=0.45]{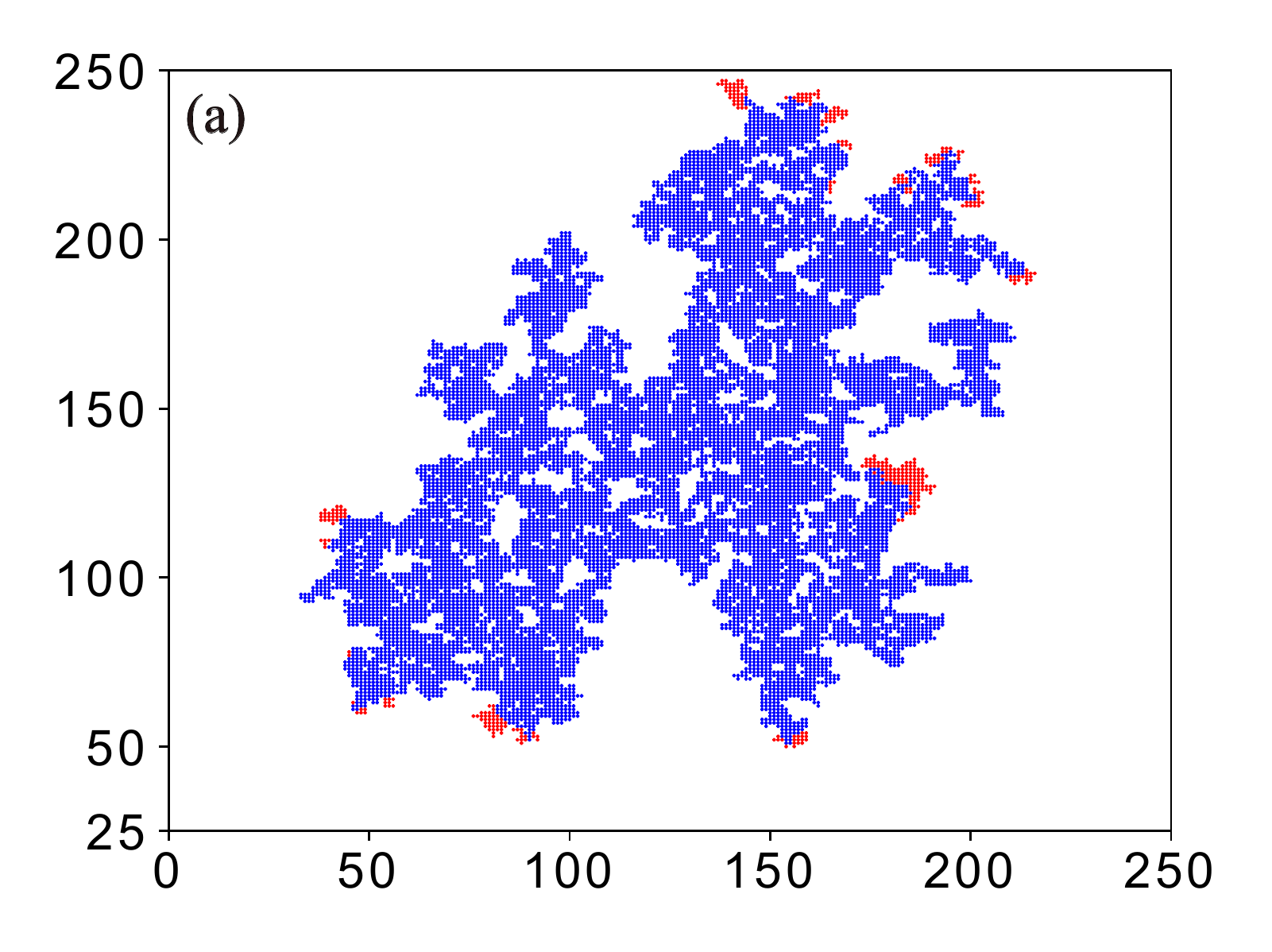}
         \includegraphics[scale=0.59]{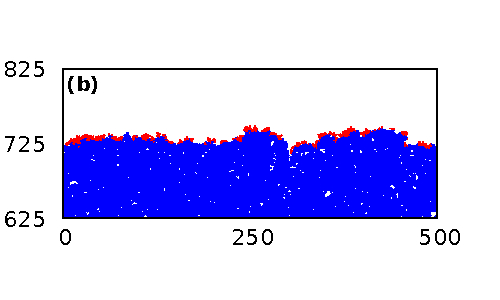}
\caption{Snapshots of realizations of the CE process with different initial conditions,  with blue representing sites occupied by the predators, and red, by prey. (a) for $p=0.495$ on a 
	$250\times250$ square lattice with a point initial condition (b) a close-up view of the \emph{front} with a line 
	initial condition at $p=0.65$ for $L=500$ at time $500$. }
\end{figure}

\section{Definition of the model}

CE percolation on $D$-dimensional hypercubical lattices is defined as follows: each site can either be occupied by a predator 
species particle (denoted by blue color), a prey species particle (denoted by red color) or it can be vacant.  In this paper, 
we shall consider two types of initial conditions: (i) Point seeds, i.e. at time $t=0$, the origin $O$ is occupied by a blue 
particle. All the neighbors of the origin are red, and all other sites are vacant. (ii) Line seeds, i.e. an entire line
(on a lattice with periodic boundary conditions) is blue at $t=0$, and the line immediately above it is red. The evolution is a continuous 
time Markov process in which a red particle can give birth to another red particle at a vacant neighbouring site at rate $p$, 
and a blue particle can eat up a red particle at a neighbouring site at rate $1$. When a predator eats up a prey particle on a 
neighbouring site, it gives birth to a predator at the site earlier occupied by the eaten prey. Note that the preys (the reds) 
can only reproduce at neighbouring vacant sites, and predators (the blues) reproduce only by eating a prey at a nearest neighbour 
site.  Note also that neither prey nor predators can die spontaneously. If all the prey particles are eaten up, the predators 
cannot grow, and  since they do not die, the system goes into one of its many absorbing states. If there are no 
predators initially, then the model reduces to the Eden model.

In Fig. 1, we show results of a typical evolution in the CE model, starting from a point initial condition  in panel (a), 
and starting from a line initial condition  in panel (b). 

This model has been studied on $k-$ary trees with initial conditions such that the predator is located at the root and the prey occupy the $k$ daughters of the root \citep{ce0,ce1}. It was shown that the critical value for coexistence in this case is given by  
\begin{equation}   
p_{c}(k)=2k-1-2\sqrt{k^{2}-k}.
\end{equation}
Note that  in large $k$ limit, this  gives $p_{c}\sim\frac{1}{4k}$. Clearly, it is seen that the prey can survive on such trees 
even if it moves at a much slower rate than the predators and that the critical value $p_c$ goes down as $k$ increases. The CE 
model was also studied on the ladder graph where the critical rate $p_c$ is $1$ \citep{ce3}. 

The CE model shows rich behaviour on $2D$ lattices and it was numerically explored in Ref. \citep{ce2}. They showed convincingly 
that even on the square lattice, coexistence between the prey and predator is possible when the rate of spreading of prey $p$ 
is strictly less than that of the predator. This is expected as the average number of vacant neighbors per prey 
particle is greater than the corresponding number of prey neighbors of a predator. 
As already mentioned,  these authors noted that in this case, $p_c$  is  very close to $1/2$.

\section{Improved determination of critical probability $p_c$: Line seeds} 

We used Monte Carlo simulations to determine more precisely the critical value $p_c$ on the square lattice. In the first set 
of runs, we simulated the process on an $L\times L$ lattice with periodic boundary conditions along the $x-$axis and initial 
conditions such that each site on the line $y=0$ is occupied by a predator particle and each site on the line $y=1$ by a prey 
particle. With this `line' initial condition, the boundaries of the clusters are statistically stationary for translations along 
the $x$-direction, and there is greater self-averaging in the evolving clusters,  and one avoids the very slowly-relaxing  fluctuations  in point seed simulations.

\begin{figure}
    \centering
    \includegraphics[scale=0.33]{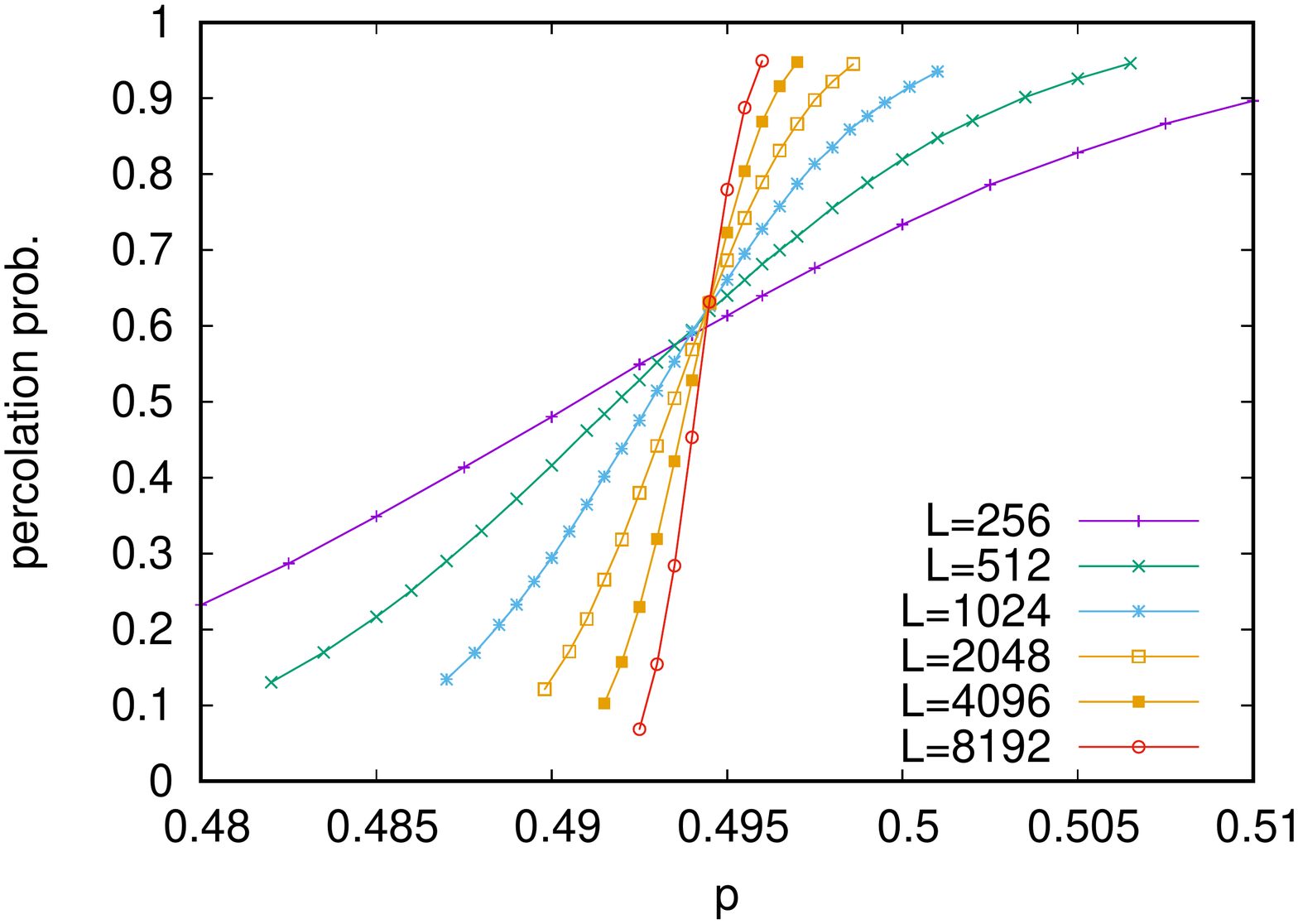}
	\caption{Survival  (`percolation') probability as a function of $p$ is plotted for $L \times L$ lattices, with 
	$L$ ranging from $256$ to $8192$. Observe the confluence of curves away from $p=0.5$.  From these simulations, 
	the critical probability is determined to be $0.4945 \pm 0.0001$.}
    \label{fig:my_2}
\end{figure}

\begin{figure}
    \centering
    \includegraphics[scale=0.33]{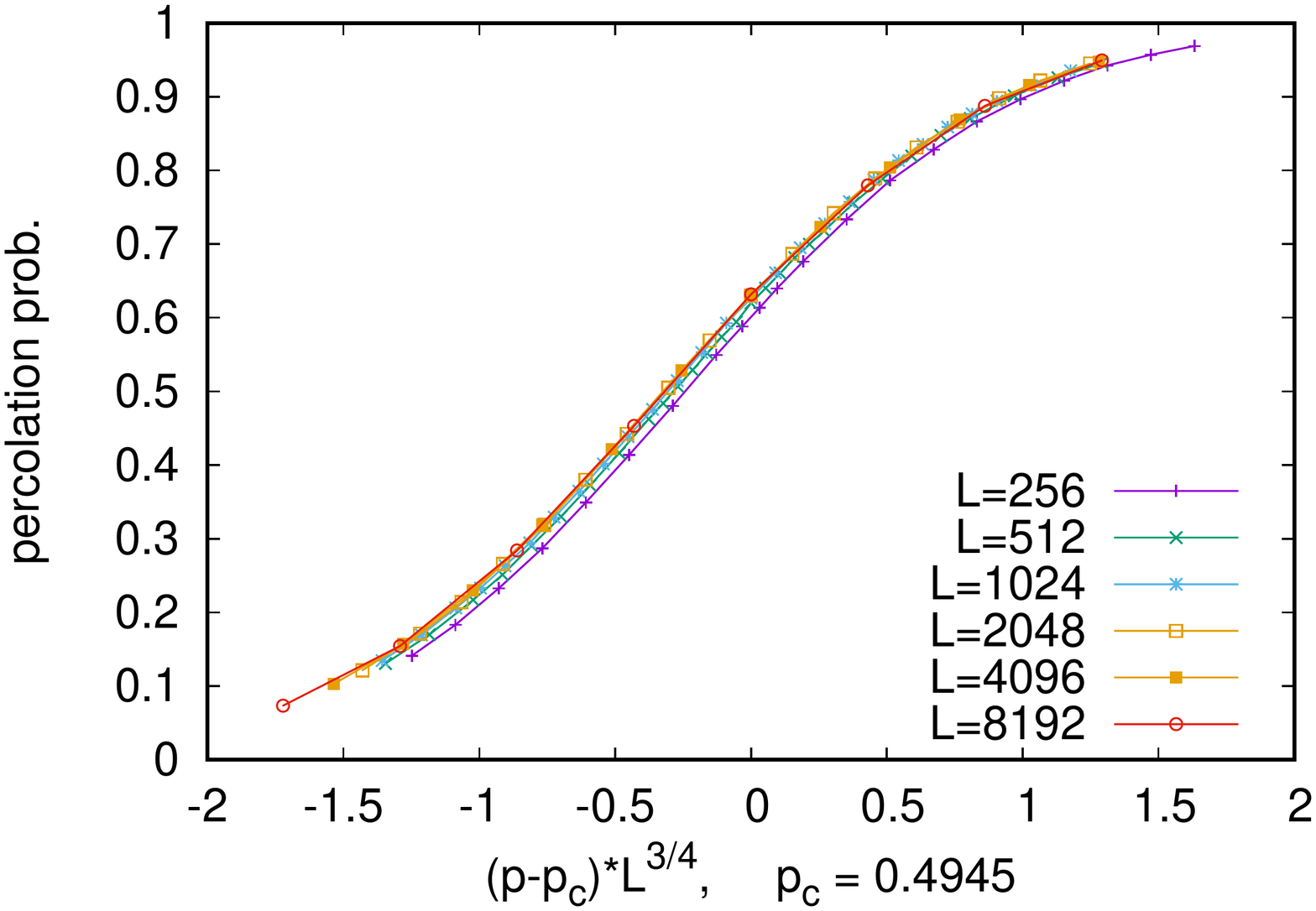}
    \caption{Scaling collapse for the survival probability is plotted against $(p-p_c) L^{1/\nu}$, with  $\nu=1.333$ and $p_c=0.4945$.}
    \label{fig:my_3}
\end{figure}

In a simulation on  a $L \times L $ lattice, we say that the prey survives (or `percolates') if any of the prey particles 
are able to reach the  top boundary $y=L$. Survival probability is then calculated by performing the simulation multiple times 
and computing the fraction of realizations in which the prey survives. We used  between $50000$ and $3\times 10^5$ 
realizations to estimate the survival probability for each value of $p$ and $L$. The plot of survival probability as function of 
$p$, for  $L$ ranging from $256$ to $8192$. is shown in Fig. 2. 

We see that the curves for different $L$ seem to meet at a common point at $p \approx 0.4945$. From finite-size scaling theory, 
this point is identified as the critical probability. From the spread of the region of intersection, we estimate the error bar 
in this estimate to be $0.0001$. We thus get estimate for the critical probability to be  $p_c = 0.4945 \pm 0.0001$. Thus, our 
simulations clearly show that $p_c$ is strictly less than $\frac{1}{2}$. 

In Figure 3, we show the results of a finite-size scaling analysis of the data.  We plot the survival probability against 
$(p-p_c) L^{1/\nu}$, and  obtain a good data collapse for the values $p_c=0.4945$ and $\nu=1.333 $. This numerical estimate of 
$\nu$ is in   excellent agreement with its value in  the standard $2D$ percolation universality class. 

\begin{figure}
    \includegraphics[scale=0.33]{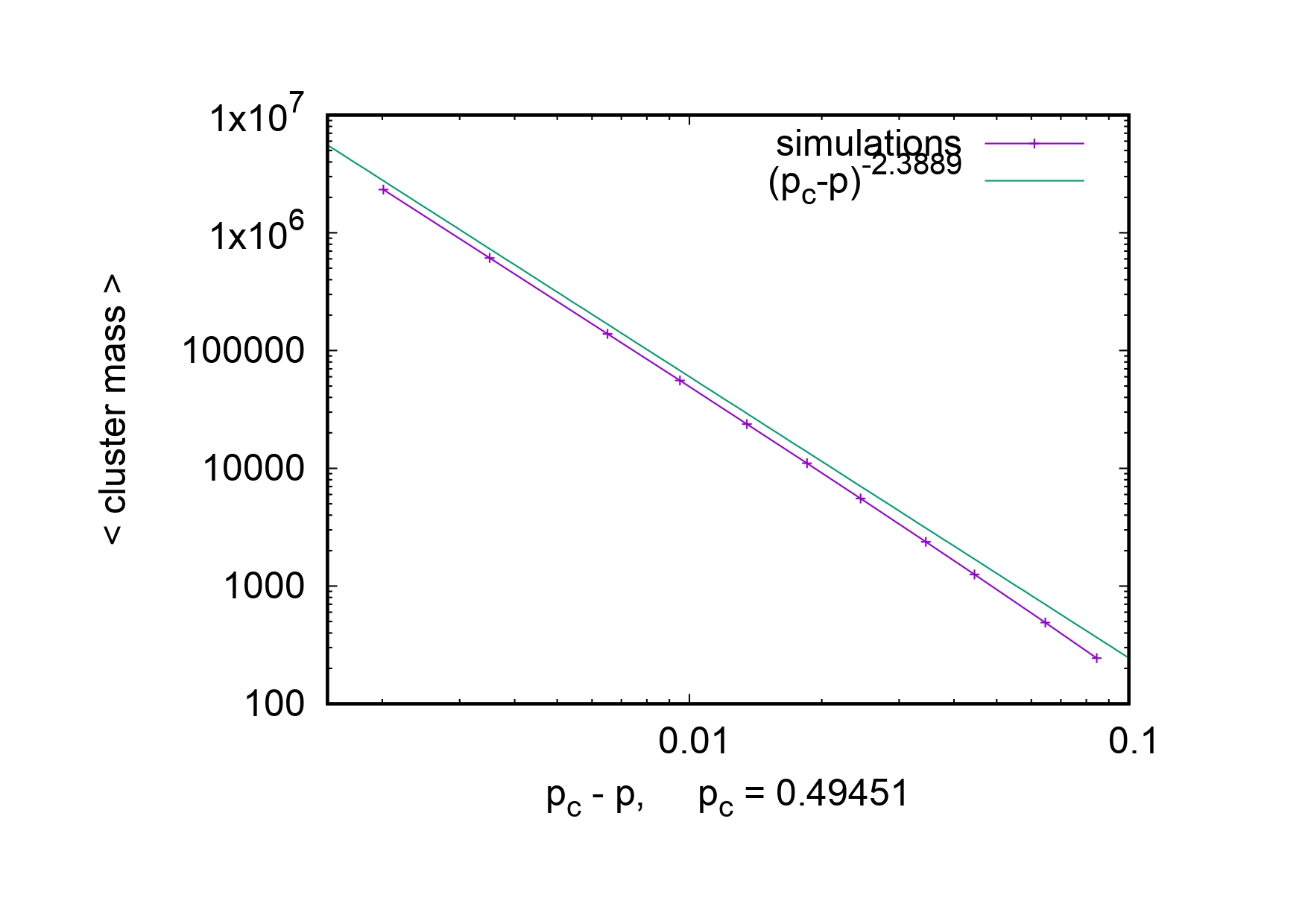}
    \caption{Average number of predators in the absorbing state as a function of $p_c-p$. Near $p_c$, the cluster size scales as 
	$(p_c-p)^{-\gamma}$ with $\gamma=43/18$, and $p_c=0.49451$. }
    \label{fig:my_3a}
\end{figure}

\section{Spreading simulations from point seeds}

Simulations starting from point initial conditions for which we observe percolation to one or more than one boundary lines would
give less precise results. Much more precise results are obtained if we observe the growth as a function of time and stop it,
before the boundaries of the lattice are reached. This is indeed the preferred strategy for directed percolation 
\citep{torre}, but it also gives very good results for standard percolation, both in low and high dimensions \cite{high-D}.
In the present simulations we used lattices of size up to $32768\times 32768$. 

In addition, one can simulate in this way also 
subcritical CE percolation, if one uses  $p$ sufficiently small so that all clusters die before they reach the lattice boundary.
Mean cluster sizes obtained in the latter way, i.e. average number of predators as $t \to \infty$ for $p < p_c$, are plotted 
as a function of $p_c-p$ in Fig. 4. 
For $p$ near $p_c$, we expect that the average cluster size would scale as $(p_c-p)^{-\gamma}$. We find a good agreement with 
$\gamma = 43/18$, which is the critical exponent for standard 2D percolation. By finding the value of $p$ that gives the best slope, 
this also allows us to refine our estimate of $p_c$ to be $0.49451 \pm 0.00002$.

\begin{figure}
     \begin{center}
	     \vglue -0.7cm
         \includegraphics[scale=0.32]{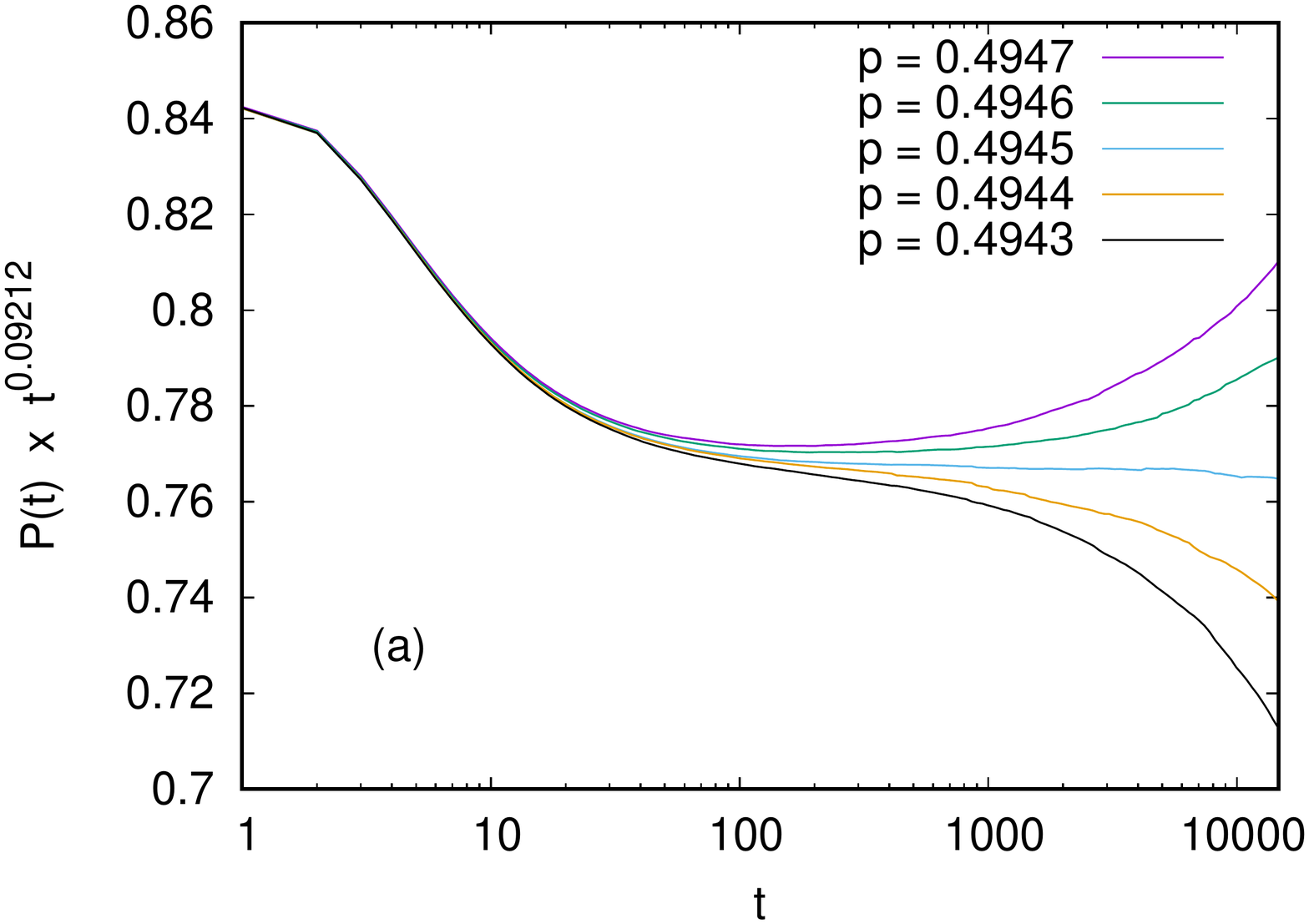}
	     \vglue -0.9cm
         \includegraphics[scale=0.32]{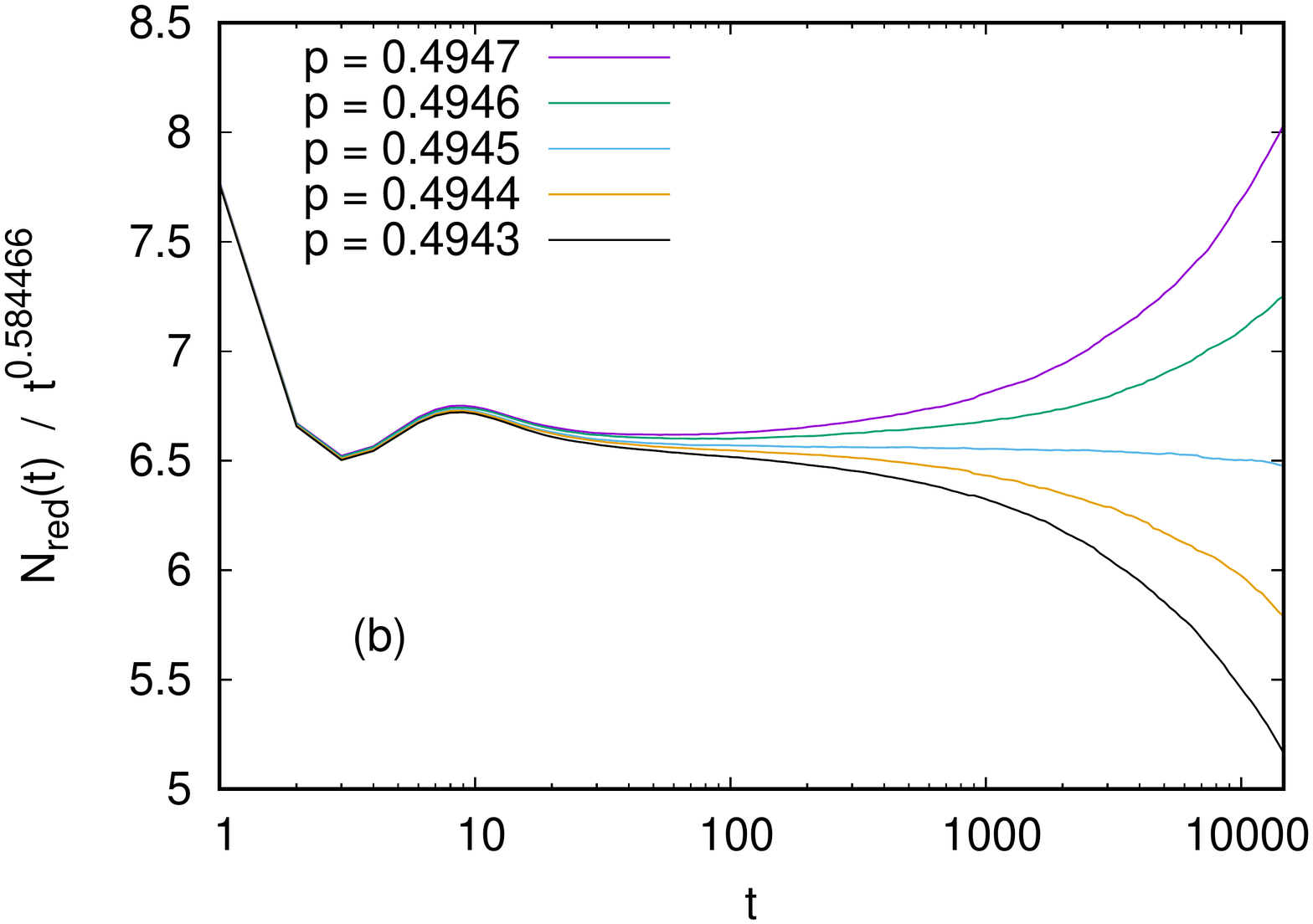}
	     \vglue -0.9cm
	 \includegraphics[scale=0.32]{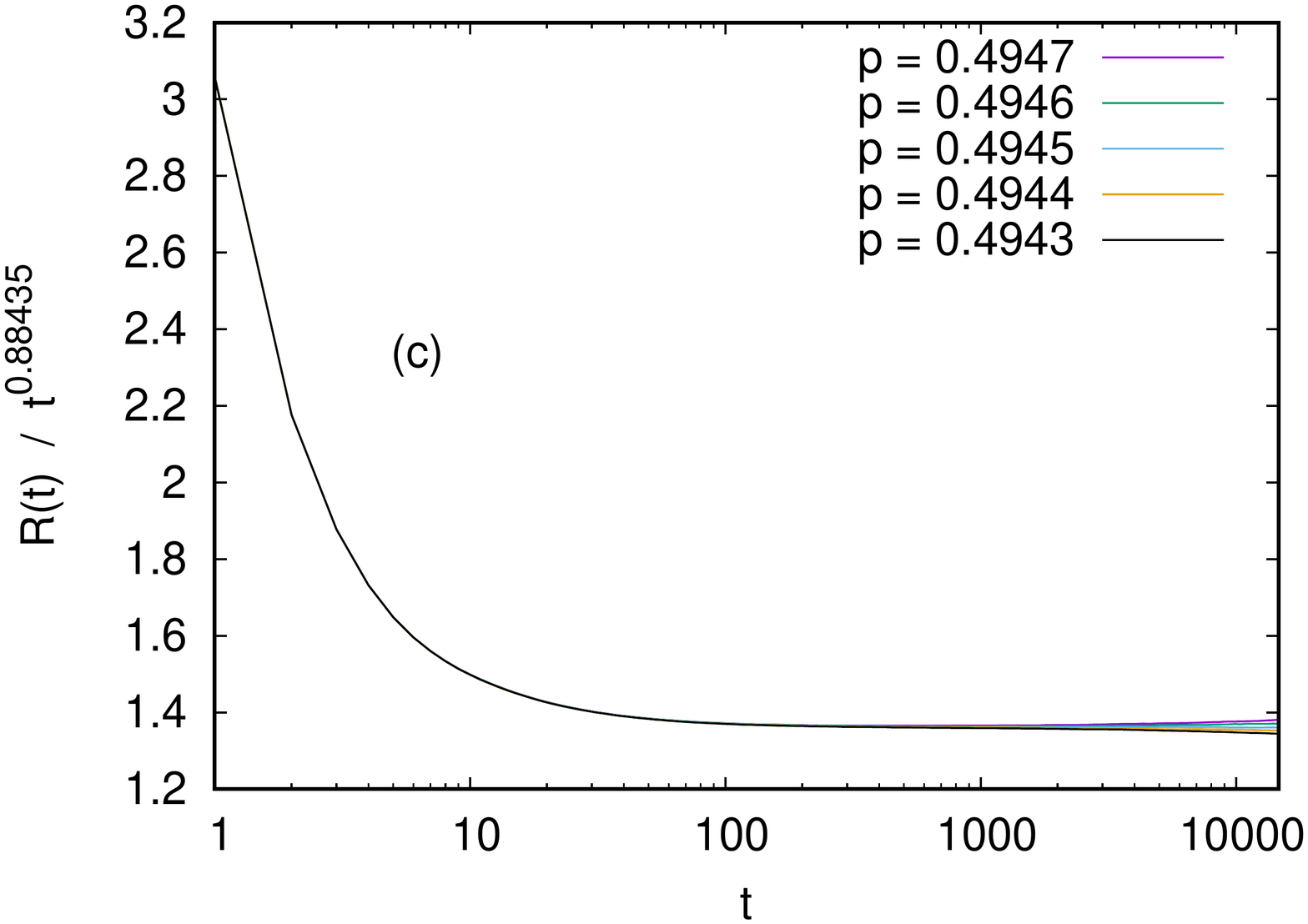}
	     \vglue -1.7cm
     \end{center}
\caption{Log-linear plots of growth observables $X=P,N_{\rm red},$ and $R$ at fixed values of $p\approx p_c$, plotted against time. For better visual discrimination, the actual variables plotted are $t^\alpha X(t)$ with $\alpha$ chosen such that the curves for 
	$p=p_c$ become flat asymptotically. Panel (a): $X(t)=P(t), \alpha = -\delta =-0.09212$; panel (b): $X(t)=N_{\rm red}(t), 
	\alpha = \mu = 0.584466$; panel (c): $X(t)=R(t),  \alpha = 1/z = 0.88435$. The values chosen for $\delta,\mu,$ and
	$z$ are indeed those for standard 2D percolation \citep{wiki}. The fact that in all three panels the curves become
	flat for $p=0.49451(1)$ shows convincingly that CE percolation is in the standard percolation universality class,
	and gives at the same time the best estimate of $p_c$.}
\end{figure}

At the critical point one expects that the prey survival probability $P(t)$, which is the probability that there exists at least
one prey particle at time $t$, scales as 
\begin{equation}
	P(t) \sim t^{-\delta},
\end{equation}
while the average number of prey particles (averaged over all runs, even those which had already died) and the average 
r.m.s. distances of prey particles from the origin scale as
\begin{equation}
	N_{\rm red}(t) \sim t^{\mu},\qquad R(t) \sim t^{1/z}.
\end{equation}
In order to test these we plot, for each of the three observables $X=P,N_{\rm red},$ and $R$ and for several values of $p$ near $p_c$,
the ratios $X(t) t^\alpha$ against $t$, with suitably chosen exponents $\alpha$. If $\alpha = -\delta, \mu,$ and $1/z$, 
respectively, we then expect the curves for $p=p_c$ to be asymptotically horizontal. Such plots are shown in Fig.~5. They 
show again that CE percolation is in the standard percolation universality class, and  linear interpolation of the slopes in  all three panels of Fig. 5 to get the $p$ value corresponding to zero slope give  our best and 
final estimate
$p=0.49451(1)$.

\section{The  discrete-time parallel-update variation}

We define below a discrete-time parallel-update variation of the CE percolation, which brings out its relationship to the 
standard percolation.  In this variation, as before, there are three states per site: unoccupied, red or blue. The time 
evolution is discrete. It is specified in terms of two parameters $p_1$ and $p_2$. In one time step,  a red site invades 
 every empty  neighboring site with probability $p_1$, and a blue site invades red  sites with probability 
$p_2$.  If a site that is unoccupied, but has $r$  red neighbors at time $t$, it  thus becomes red at time $(t+1)$ 
with probability $[ 1-( 1-p_1)^r]$, and remains unoccupied otherwise. A red site with $r'$ blue neighbors at time $t$, 
becomes blue at time $(t+1)$ with probability $[1- (1-p_2)^{r'}]$. A blue site at time $t$ remains blue at all subsequent 
times. 

\begin{figure}
    \includegraphics[scale=0.34]{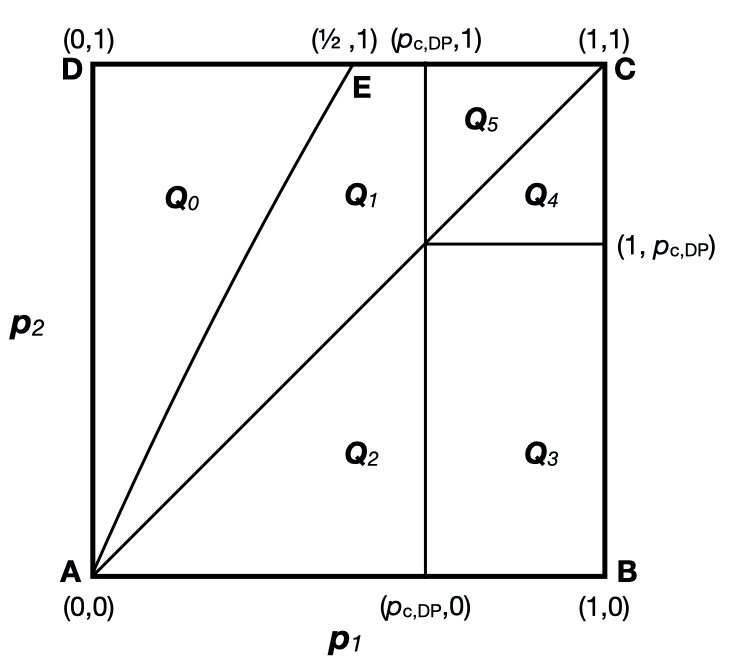}
    \caption{ A schematic depicting different phases in the parameter space $(p_1, p_2)$ of the generalized Chase-Escape model with parallel updates. The parameter space is divided into  six  phases, where $Q_0$ denotes the absorbing phase, and $Q_1$ to $Q_5$ denote the five different active phases discussed text.}
\end{figure}

\begin{figure*}
    \includegraphics[scale=0.8]{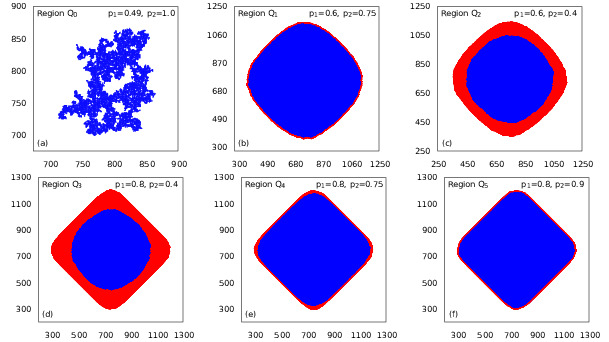}
    \caption{ Snapshots of realizations of the generalized discrete time-parallel update CE process with point initial conditions for the six different phases at selected representative points.}
\end{figure*}

The parameter space here is $(p_1,p_2)$, both of which lie in the interval $[0,1]$. This is shown as a square $ABCD$ in 
Fig. 6. It is easy to determine the behavior of the model along the boundaries of this square, say starting with point 
initial conditions. Also, the continuous time model studied in previous sections is recovered by setting 
$p_1= p \delta t, p_2 = \delta t$, in the limit $\delta t$ tends to zero. 
Along the line $AD$,  we have $ p_1=0$, and the system will reach an absorbing state, with exactly five blue sites, for 
all $p_2 >0$. Along the  $ p_2=0$ line $AB$, the blue cluster does not grow, and the red cluster grows as in the 
discrete-time Eden model (also called the Richardson model \cite{richardson}). The radius of the red cluster 
increases linearly with time. For $p_1=1$ ( the line $BC$), the red cluster grows deterministically, as a diamond, with 
the maximum velocity $1$. All the perimeter sites of the growing blue cluster are red, and it grows as the Richardson 
model with growth parameter $p_2$.

Along the line $CD$, with $p_2=1$, it is easily seen that at any time,  a red site  has always  at least one blue neighbor. 
Thus it survives for only one unit of time. We assign to each bond \emph{i.i.d.}  random variables `occupied' with probability 
$p_1$, and 'unoccupied' with probability $(1-p_1)$.  We make a red site at time $t$ blue at time $(t+1)$ iff  it had  
a blue neighbor at time $t$ connected by an occupied bond.  Then clearly, the stochastic evolution of this model is as 
defined  above.  Also, the probability that the blue sites eventually constitute a finite cluster  ${\cal C}$ is 
the same as the probability of that cluster ${\cal C}$ in standard bond percolation with occupied bond density  $p_2$.  

Since the percolation probability for bond percolation on the square lattice is well-known to be $1/2$, we see that the 
point $E$, which marks the end of the absorbing phase in Fig.~6 has coordinates $(1/2,1)$, and the critical exponents 
at this point are the same as in the standard bond percolation. If we assume that  critical behavior is same all along 
the line $AE$, as seems reasonable, then the universality class of CE-percolation is the same as standard percolation. 
Also, the line $AE$ has average slope $2$, and the fact our estimated $p_c$ in the continuum case was near $1/2$ 
indicates that its  slope differs from $2$ only by small amount all along the curve. 

The region in the $(p_1,p_2)$ plane to the right of line $AE$ is the region where prey can survive indefinitely with a 
non-zero probability. Using the result of Durrett and Liggett about the existence of linear segments on the limiting 
front shape \cite{durrett-liggett}, we can divide this region into five sub-regions, denoted by $Q_1$ to $Q_5$ in Fig.~6. These are defined as follows: In $Q_1$, both red and blue fronts are pinned together in all directions. The common velocity of the front is weakly direction dependent, but no front has a linear part. In $Q_2$, the fronts are fully depinned, and  the red front moves faster than  the blue  one in all directions. In $Q_3$, the fronts 
are  also fully depinned, but for $p_1 > p_{c,DP}$, where $ p_{c,DP}$ is the critical probability for directed percolation, the 
red front has  four linear segments, 
where the front velocity, in the $L_1$ norm, reaches the maximum possible value $1$. For the square lattice, $ p_{c,DP} $ is known rather accurately: $ p_{c,DP} \approx 0.644700185(5)$ \cite{jensen}. In region $Q_4$, 
the blue front also reaches the maximum possible velocity in some directions, and then in these directions, the red 
and blue fronts are again pinned together, and in $Q_5$, the fronts are fully pinned again. Clearly, the boundary between $Q_2$ and $Q_3$ is a vertical line, with equation $p_1 =  p_{c,DP}$, whereas $Q_3$ and $Q_4$ are separated by the line $p_2= p_{c,DP}$.
The line $p_1=p_2$ forms the boundary between  $Q_1$ and  $Q_2$, and also between   $Q_4$ and $Q_5$.

\section{LOWER BOUND TO THE CLUSTER SIZE
IN THE ABSORBING STATE}

Consider the CE percolation on the $D$-dimensional hypercubical lattice starting from point seed initial conditions (a single 
predator at the origin, surrounded by $2D$ prey particles on its neighbouring sites). We consider the case  $p<p_c$, when the 
system eventually reaches one of its absorbing states and we are left with a cluster of predators.

\begin{figure}
    \includegraphics[scale=0.34]{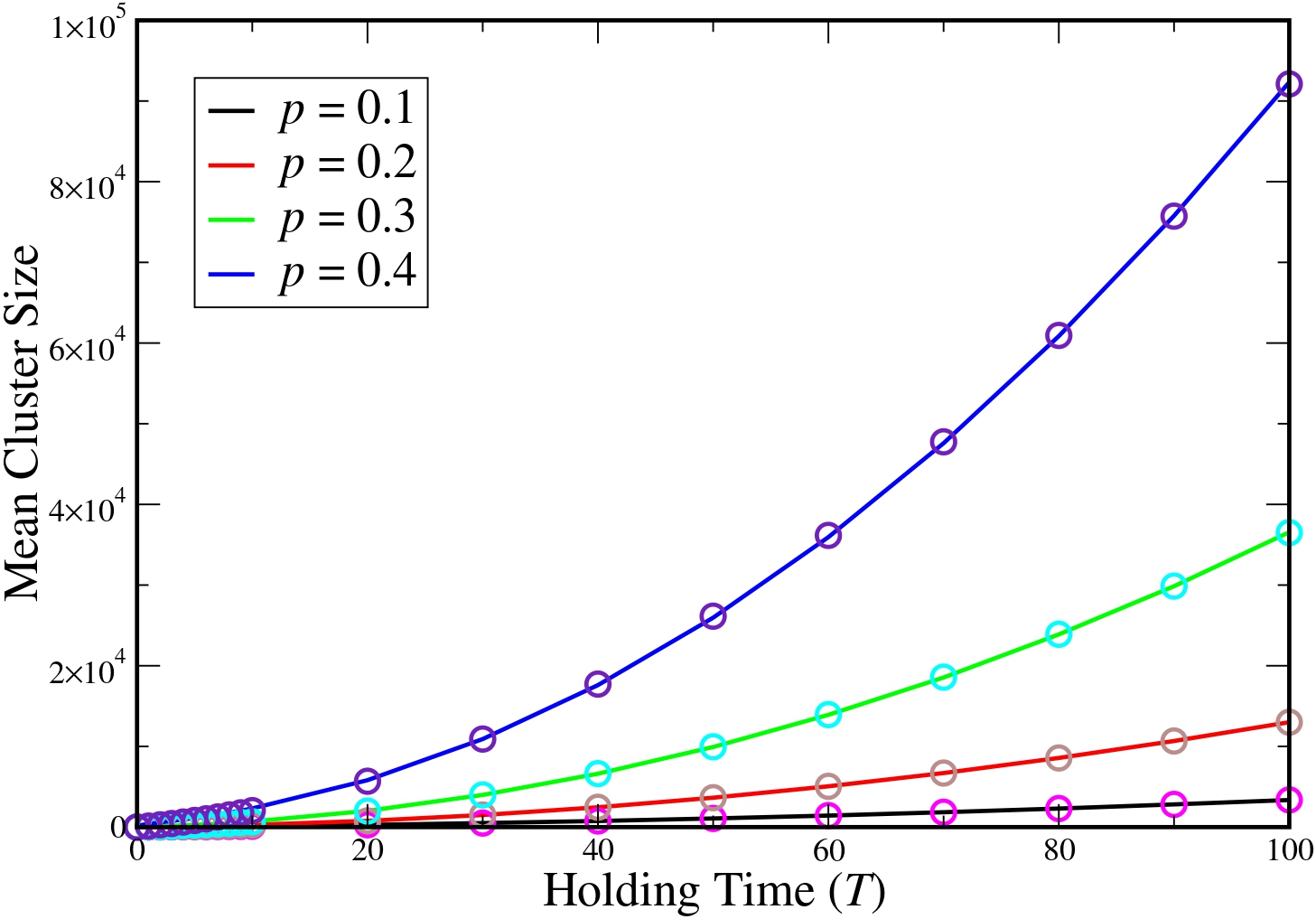}
    \caption{Mean cluster size of predators in the absorbing state of CE percolation as a function of holding times $T$, 
	for $p=0.1,0.2,0.3$ and $0.4$. To each plot, we fit a function $Y=aT^2 + bT$ and obtain an excellent fit.}
    \label{fig:my_3b}
\end{figure}

\begin{figure}
    \includegraphics[scale=0.34]{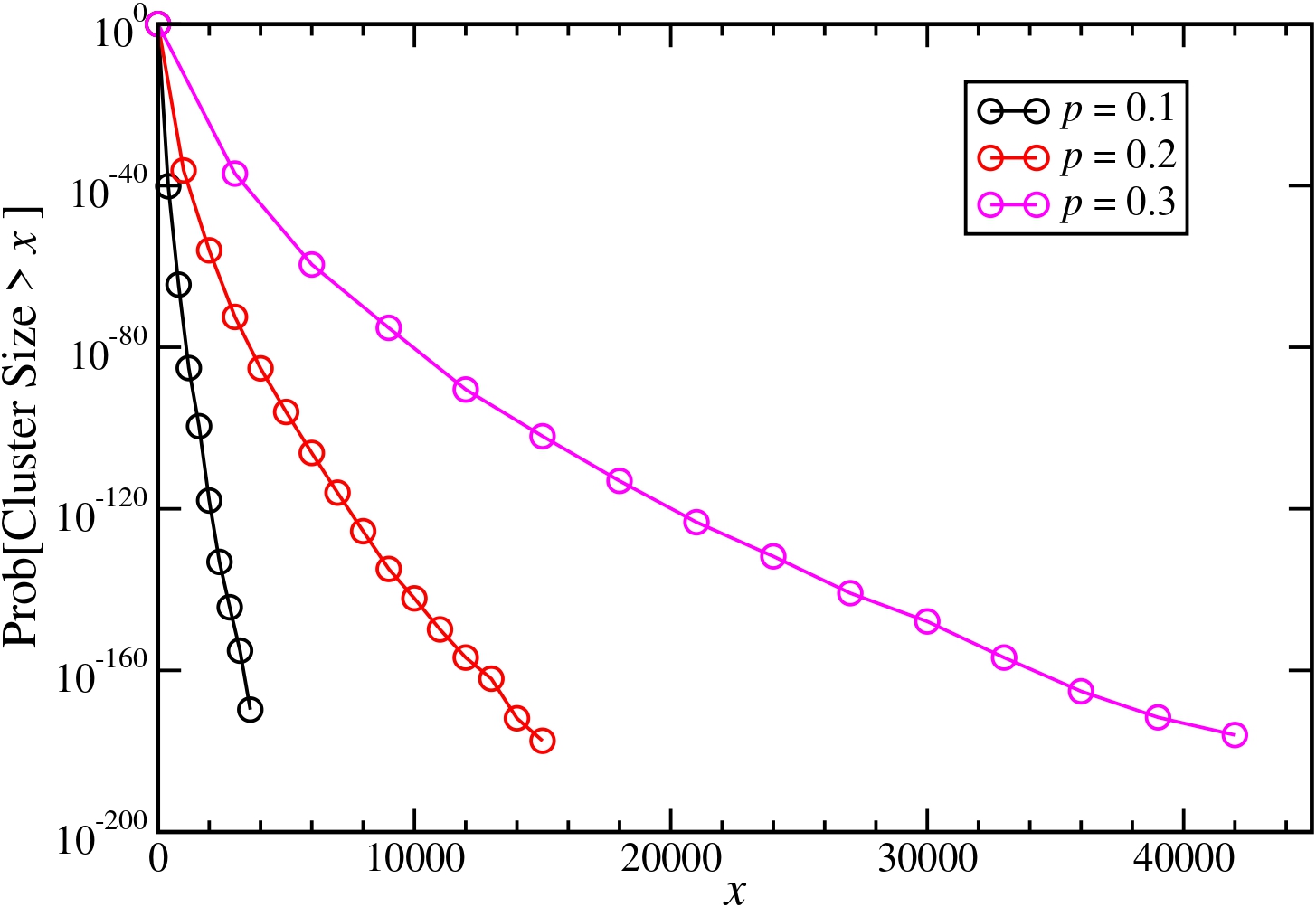}
    \caption{Cumulative probability as defined in Eq.~(4) for $p=0.1, 0.2$ and $0.3$, computed using the rare event 
	sampling technique as described in the text. The $x$-axis is linear, where as the $y$-axis is plotted on
	logarithmic scale. The plot shows a clear deviation from exponential behaviour, and provides evidence in favour 
	of our stretched exponential lower bound.}
    \label{fig:my_4}
\end{figure}

Let $\textrm{CumProb}(s)$ be the cumulative probability that, in the absorbing state, the number of predators is greater than $s$. 
It is clear that this probability will be bounded from below by the probability that the prey population reaches $s$ 
before the predator present at the origin eats up its first prey. The probability that the predator at the origin does 
not eat any of its adjacent prey up to time $T$ is $\exp(-2DT)$, but in that much time, the number of prey particles 
grows to order of $(pT)^D$. It is clear that for the prey population to reach size $s$, the time required would be of 
order $s^{1/D}/p$. This immediately gives us
\begin{equation}
\textrm{CumProb}(s)\geq C_{1}\exp\left[-\frac{C_{2}s^{1/D}}{p}\right]
\end{equation}
where $C_1$ and $C_2$ are constants, which may be chosen to be independent of $p$. This is in contrast to standard 
percolation, where $\textrm{CumProb}(s)$ decreases exponentially with $s$ for large $s$, for values of $p$ below the critical threshold.

To accurately capture the tail of the subcritical cluster size distribution in numerical simulations, we adopt a 
rare event sampling approach, where we delay any action of the first predator by a holding time $T$, for various 
values of $T$, and let the prey grow freely till then, without facing any predation. After time $T$, the process 
follows its usual evolution and we note down the final cluster size and assign it a probability weight of $\exp[-4T]$. 
In Fig.~6, we plot the mean size of the final predator cluster as a function of the holding time $T$, for $p=0.1, 
0.2, 0.3$ and $0.4$, and to these plots, we fit a function $y = aT^2 + bT$. The fit shows very good agreement with our 
data. For $p=0.2$, the best values of $a$ and $b$ are $1.13$ and $15.98$ respectively, whereas for $p=0.4$, we have 
$a=7.96$ and $b=124.68$. This suggests that clear indications of the stretched exponential tail would come at large 
cluster sizes (corresponding to large holding times). In Fig. 9, we plot the full cumulative distribution, as defined 
in Eq.~(4), of the cluster size distribution for $p=0.1, 0.2$ and $0.3$ using the importance sampling method. To obtain 
the tail behaviour and sample the full distribution, we plot the weighted distribution obtained by merging the cluster 
size distributions for holding times $T$ ranging from $1$ to $10$ in spaces of $1$ unit time, and $20$ to $100$, in 
steps of $10$ units of time. For each holding time, we perform $20000$ realizations to obtain the cluster size 
distribution. The plot shows a clear deviation from exponential behaviour, and provides evidence in favor of our 
stretched exponential lower bound.

This stretched exponential decay seems to be inconsistent with the exponential  decay for standard percolation. If 
CE percolation is in the universality class of the usual $2D$ percolation problem, then not only the critical exponents, 
but also the scaling functions in the critical region should be universal. In the  standard percolation, the probability distribution of cluster size $s$ has the scaling form
\begin{equation}
\textrm{Prob}(s) \sim s^{-\tau} f( s \epsilon^{\phi}),
\end{equation}
as $ \epsilon \to 0$, with 
$s \epsilon^{\phi}$ fixed. Here, $\epsilon = p_c -p$, and $f(x)$ is known to decay as  $\exp(-x)$,  for large $x$.
However, we can have 
\begin{equation}
 {\rm Prob}(s)  \sim s^{-\tau} \exp (- B_1 s \epsilon^{\phi}) + \exp( - B_2 s^{1/D}).
\end{equation}
  Clearly, the 
stretched exponential decay of cluster size distribution dominates only  for $s > \epsilon^{-\phi D/(D-1)}$, but then the 
scaling variable $x$ in the scaling function would tend to infinity. Thus, the asymptotic stretched exponential decay of 
${\rm Prob}(s)$ for large $s$  is consistent with the exponential decay of the scaling function $f(x)$ for large $x$. 

\section{Direction dependence of survival probability}

The qualitative reason why the faster predators can have the same front velocity as the prey in the regime  $p\leq p_c$ 
(and, as we shall see later, also for $p_c < p \leq 1$) is clearly due to the fact that the predators can only move 
along the sites occupied by the prey. But the paths created by the prey along different directions are qualitatively 
different.  This is well known in the Eden model, which is just the model where the prey spreads without interference of 
predators \citep{dhar-eden,wolf}. The first hints that the spreading velocities in the 2D Eden model are different along
the diagonal and parallel to the axes were obtained in \citep{plischke}, while the most precise simulations, by Alm and 
Deijfen, show that the velocity along the diagonal is smaller than the velocity along one of the axes by roughly 
$1.3\%$ \citep{fpp2d}.

Also, in most of the regions of the $(p_1,p_2)$ plane of the discrete-time problem, the velocities of front 
are direction dependent. The rotational invariance is expected to emerge only at the critical point, in the scaling limit. 
Clearly, the geometrical structures  of paths with different average orientations are somewhat different. 
If the front velocity in different directions is different, even the critical thresholds could be different 
in different directions. We recall that in the well-studied case of directed percolation, there is a orientation 
dependence of the critical threshold $p_c(\theta)$, and infinite directed paths in the direction $\theta$ appear only 
if $p > p_c(\theta)$.

Since the isotropic scaling limit is an important prediction of the conformal field theory that gives the exact 
values of critical exponents of the percolation theory, it is useful to have a direct test of the restoration of 
isotropy, {\em at the critical point, in the scaling limit}. Rigorous proofs of this are not easy, but at least for 
bond percolation and several other models such a proof was given recently in Ref.~\citep{duminil}. In the following we shall present two different sets of simulations for directly  verifying this  prediction.

\subsection{Line seed spreading from a tilted line}

 \begin{figure}
     \centering
     \includegraphics[scale=0.34]{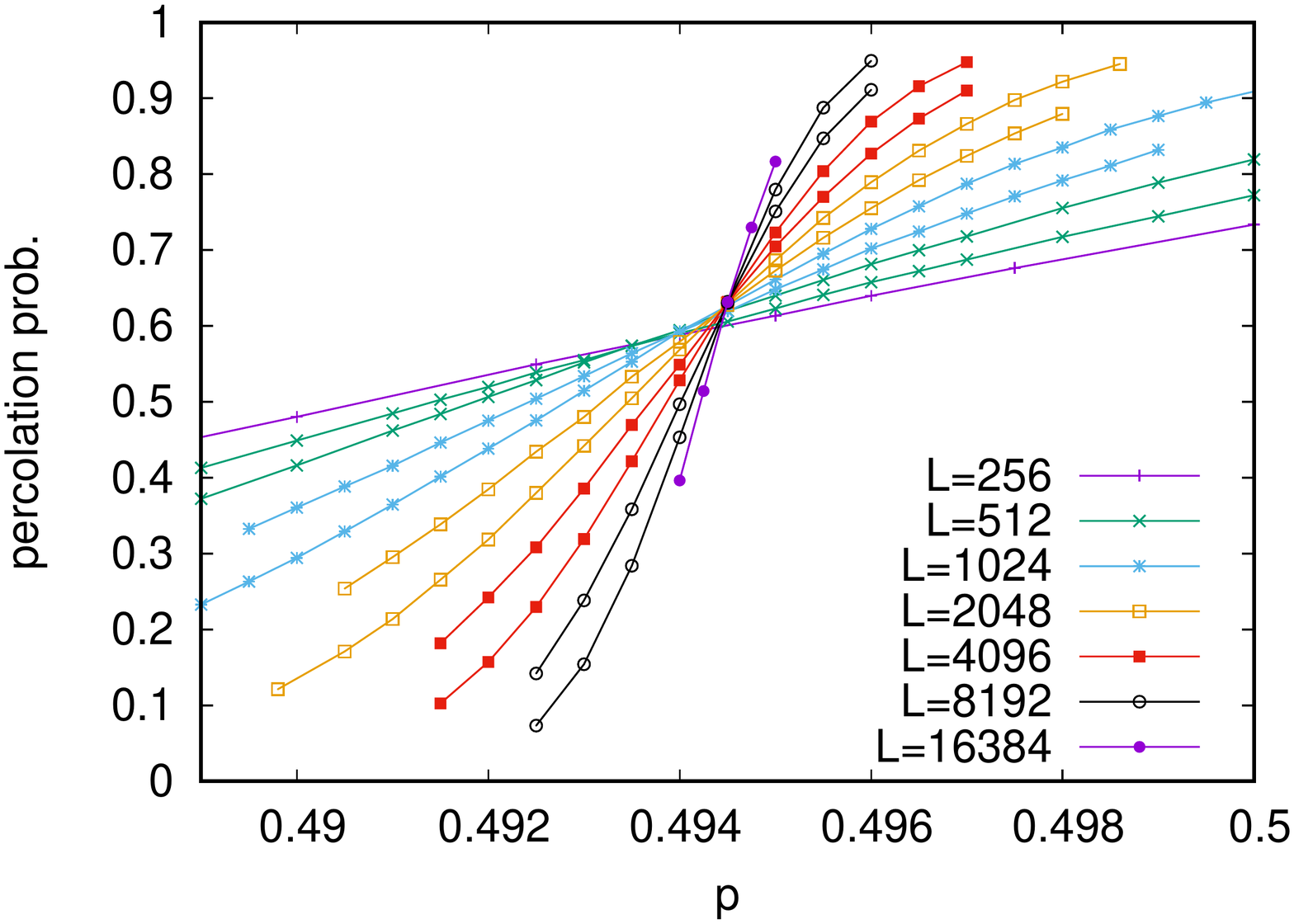}
     \caption{Survival  (`percolation') probability as a function of $p$ is plotted for $L \times L$ lattices, both 
	 with the original orientation (data are from Fig.~2) and for lattices tilted by 45 degrees so that the spreading 
	 occurs effectively along one of the two diagonals of the untilted lattice. For each pair of curves with the same 
	 color, the shorter and less steep one is for the tilted lattice.}
     \label{fig:8}
 \end{figure}

In Sec.~III we had measured $p_c$ by observing the spreading from a line seed that was oriented parallel to the x-axis. In 
complete analogy, we can also use a line seed parallel to one of the two diagonals. Equivalently, we can replace the 
neighborhood $\{(x\pm 1,y),(x,y\pm 1)\}$ of any site $(x,y)$ by the neighborhood $\{(x\pm 1,y\pm -1),(x\pm -1,y\pm 1)\}$, 
and start again from an initial condition where even sites on the x-axis are occupied by predators, and odd sites on the 
line above by prey. Again we use periodic boundary conditions laterally, and measure the fraction of percolating runs where
we say that a run percolates if at least one prey particle reaches the top line of an $L\times L$ lattice. Notice that 
now only half of all lattice sites (where $x+y$ is even) participate in the dynamics, thus we can simulate somewhat larger
lattices than in Sec. III. 

 Results are shown in Fig.~10, together with those from Fig.~2. We see that both sets of curves cross (in the limit of large 
$L$) at the same value of $p$. Thus $p_c$ is indeed independent of the orientation. We can also try a data collapse in the 
way of Fig.3. When doing this, we just have to take into account that the process now takes effectively place on a tilted
square lattice with lattice constant being $\sqrt{2}$, thus we should use on the x-axis $(p-p_c) [L/\sqrt{2}]^{3/4}$. This is shown in  Fig.~11. We see that the scaling curves in the two different directions show excellent collapse, verifying that the scaling function is independent of direction.

\begin{figure}
    \centering
    \includegraphics[scale=0.34]{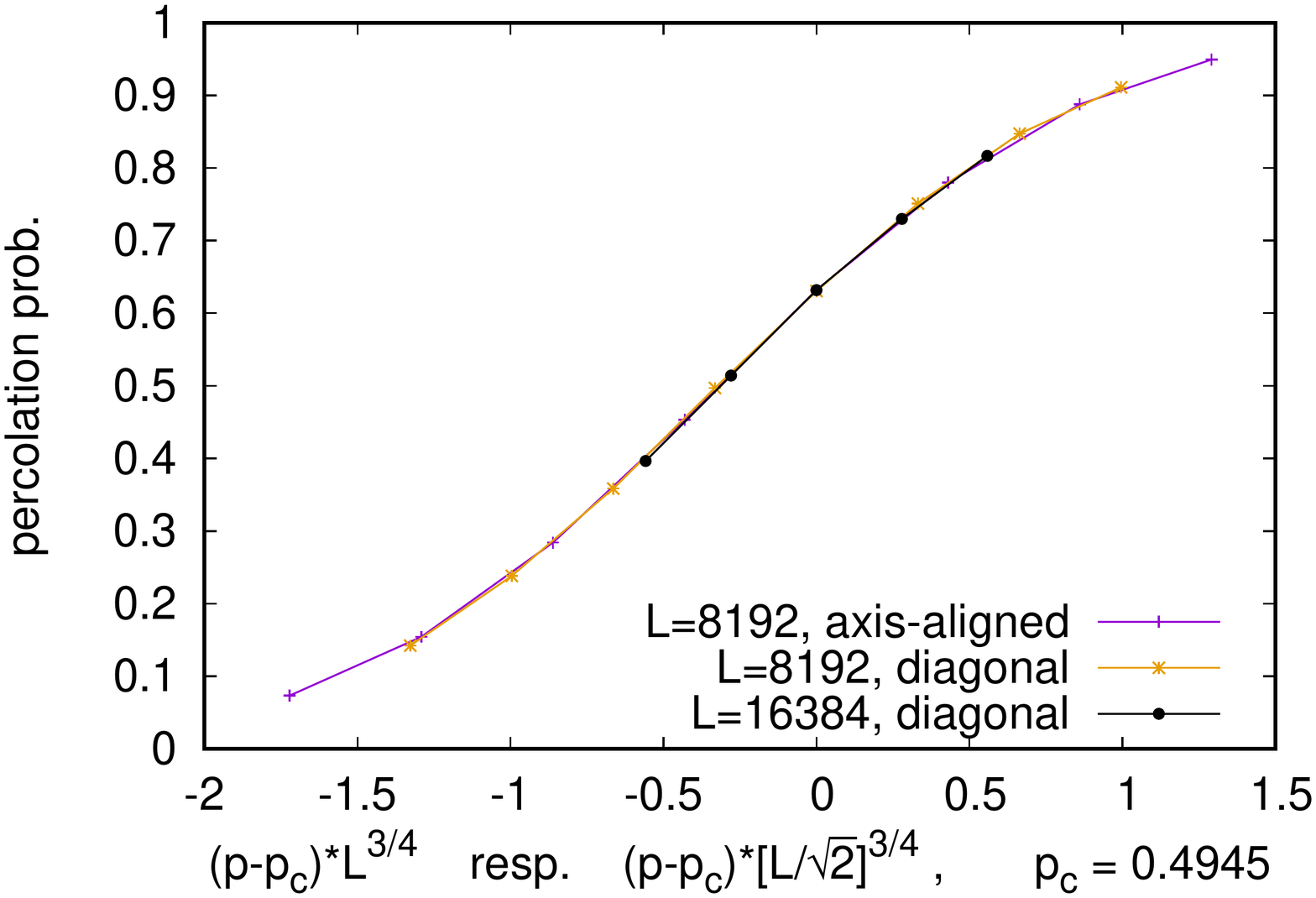}
    \caption{Testing the rotational invariance at the critical point: The scaling function for the survival probability as a function of the normalized scaling variable  $(p-p_c)L^{3/4}$ for propagation along the axis  shown for two values of $L$, collapses to that calculated along the diagonal, if the diagonal value of the lattice size $L$ ( which is an  integer, measured in lattice units) is scaled by $1/\sqrt{2}$.  Curves for other $L$ are omitted for visual clarity.}.
\end{figure}

\subsection{Anisotropy of fourth order moments in clusters grown from point seeds}

\begin{figure}
    \centering
    \includegraphics[scale=0.34]{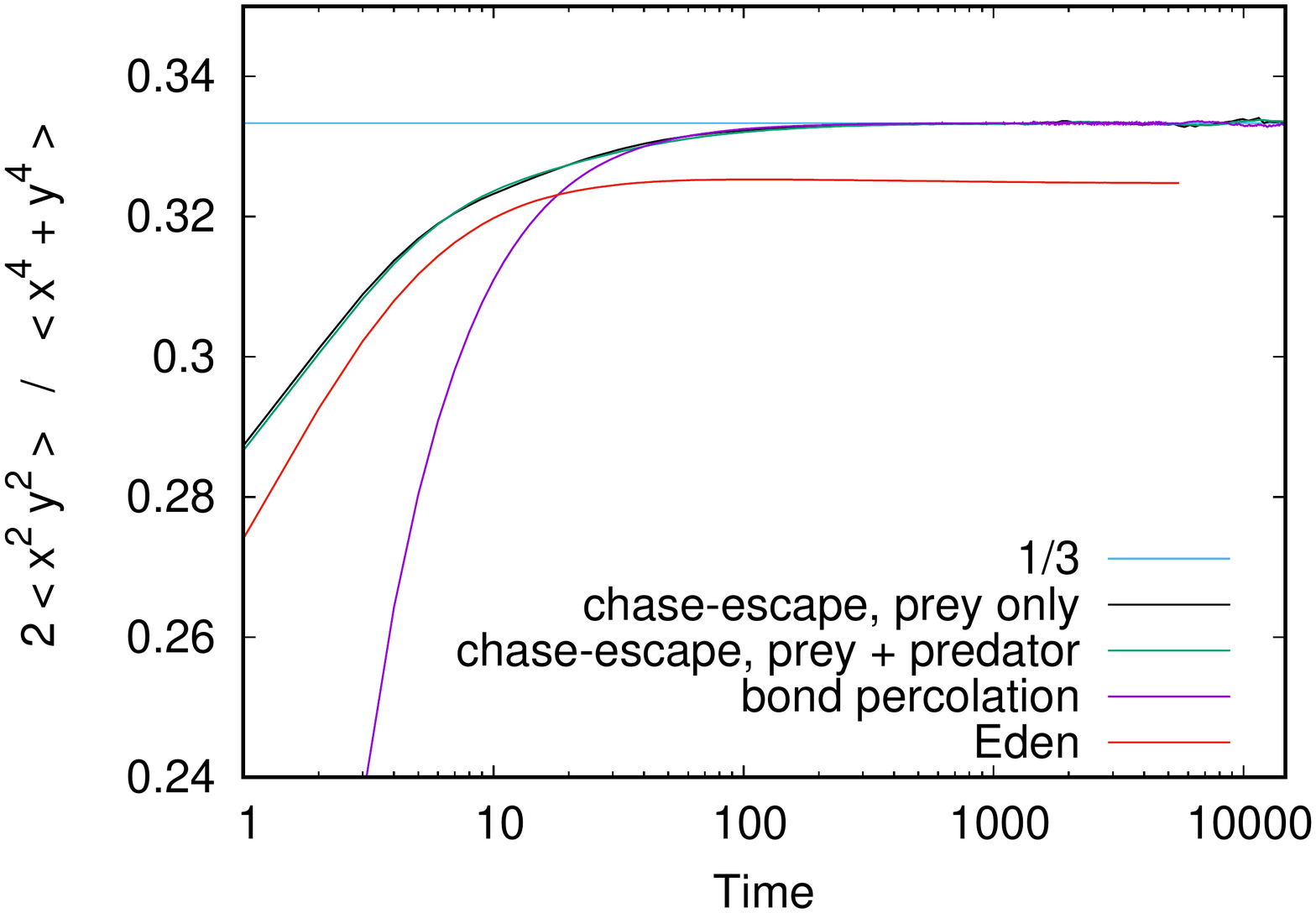}
    \caption{Log-linear plot of the ratio $2\langle x^2 y^2\rangle /\langle x^4+ y^4\rangle$ of fourth moments, for clusters
	grown from point seeds. For isotropic clusters this ratio is $1/3$, for any radial distribution. For the Eden model
	this ratio is clearly smaller than $1/3$, showing again its anisotropy. For the other three cases (bond percolation,
	prey in critical CE percolation, and prey + predators in critical CE percolation) isotropy is violated for small 
	times, but is restored in the limit $t\to\infty$}.
    \label{fig:9}
\end{figure}

For any radially symmetric distribution $\rho(x,y)$ one has 
\begin{equation}
	\int dx dy \rho(x,y) x^2 y^2  = \frac{1}{3} \int dx dy \rho(x,y) x^4 \;.
\end{equation}
Therefore, if there were no lattice anisotropy, clusters grown from standard 2D percolation, from CE percolation, and from 
the Eden model would all satisfy
\begin{equation}
	2 \langle x^2 y^2 \rangle = \frac{1}{3}\langle x^4+ y^4\rangle \;.     \label{isotrop}
\end{equation}
Any deviation from this must be a consequence of lattice anisotropy. For short times, this anisotropy is of course 
unavoidable, but exactly at the critical point, one expects that isotropy is restored at large length scales and Eq.~(\ref{isotrop}) 
holds in the limit $t\to\infty$. 

In Fig.~12 we plotted the ratio $2\langle x^2 y^2\rangle /\langle x^4+ y^4\rangle$ against time $t$, for several models:
The Eden model (with $p$ equal to $p_c$ of CE percolation), critical bond percolation, and CE percolation. For the 
latter we show both moments of prey and moments of all (prey + predator) particles. The latter two are indistinguishable
on this plot. The Eden model is clearly anisotropic (with $2\langle x^2 y^2\rangle /\langle x^4+ y^4\rangle \to 0.3246(1)$
for $t\to\infty$), but the others are clearly isotropic. The rate of convergence towards $1/3$ seems to be different 
from that in  standard percolation, but this could also be a consequence of the fact that we needed to start from a 
larger seed in the latter.

\section{Dynamics of the fronts for $p> p_c$}

A picture of the interface for $p=0.65$ was shown in Fig.~1(b).  We can see that the red sites form a rather large number 
of disconnected clusters. The front does become smoother for larger $p$. There are several possibilities to define 
such fronts and their positions precisely. For our discussion it is convenient to work with 
single-valued height functions $h_{\rm red}(x,t)$ and $h_{\rm blue}(x,t)$, called the red and blue front heights at time $t$, that 
specify the largest $y-$coordinate among all sites ever reached with that value of $x-$coordinate by that color up to time 
$t$. We will study here the mean values and variances of $h_{\rm red}(x,t)$ and $h_{\rm blue}(x,t)$, and their dependence on $t$. 
Note that if for some $x$, there is no red site at time $t$, then that particular value of $x$ is disregarded during the 
computation of the height statistics. Alternatively, we can -- instead of defining explicitly the prey front and 
the corresponding $h_{\rm red}(x,t)$ -- define only the predator front and the front of all non-white sites, with heights
$h_{\rm blue}(x,t)$ and $h_{\rm tot}(x,t)$ respectively. In this way we do not have to discard any values of $x$. It is easy to see that for $p>1$, 
the mean distance $w(t)$ between the prey and 
predator fronts increases linearly with time,  as does also the average of $\delta h(x,t) = h_{\rm tot}(x,t)-h_{\rm blue}(x,t)$ 
and the density of red sites $\rho_{\rm red}(t)$ per unit $x$. For $p<p_c$, the system goes into one of its absorbing 
states and for large $t$, $\rho_{\rm red}(t)$ and $\delta h(x,t)$ are exactly zero. \\

We studied the dynamics of the prey and predator fronts by numerical simulations on a lattices of dimensions up to $65536 \times 32678$ 
with periodic boundary conditions along the $x-$axis, unless explicitly stated otherwise. For $p_c<p<1$, we find that the 
centre of mass of the prey and predator fronts move together at the same velocity. In this regime, the two fronts are 
\emph{pinned} together and at $p=1$, a \emph{depinning transition} \citep{depin1,depin2} occurs.

\begin{figure}
    \centering
    \includegraphics[scale=0.34]{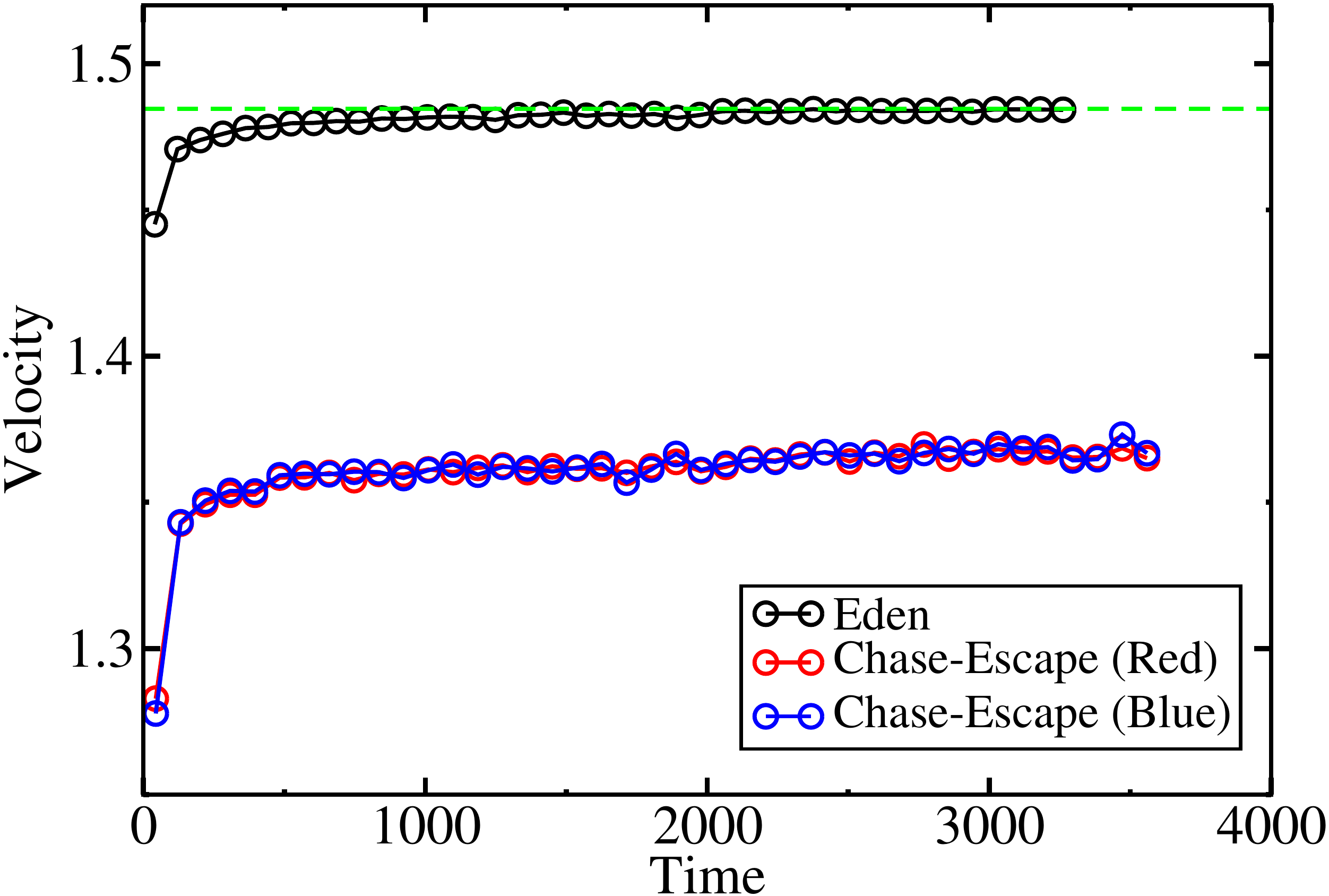}
	\caption{Average velocities  for $p=0.6$ of the three fronts over a sliding window of 100 units as a function of 
	time, to check for convergence of velocity. The dashed green line depicts the Eden velocity from the simulations of 
	Alm and Deijfen \citep{fpp2d}.}
    \label{fig:my_dfg}
\end{figure}

In Fig.~13, we plot the average velocities of the prey and predator fronts (measured over a sliding window of 
size 100 units) at $p=0.6$, and also of the Eden front using the same value of $p$. The dashed green line is the velocity of 
the Eden front as obtained by Alm and Deijfen \citep{fpp2d}. Our simulation of the Eden front is in agreement with their 
result. It is clear that the front velocity in CE is strictly less than what it would be if  predators were absent. \\

This is somewhat surprising. It was shown by Owen and Lewis \citep{predation} in a rather general continuum model of 
predator-prey dynamics that  the prey front velocity is unaffected by the presence of a predator, except if some 
special conditions are satisfied. The dynamics of Owen and Lewis model is described by the equations  
\begin{equation}
\begin{aligned}
\frac{\partial u}{\partial t} &= \epsilon D \frac{\partial^2{u}}{\partial{x^2}} + ru f(u) - \phi v h(u)    \\
 \frac{\partial v}{\partial t} &= D\frac{\partial^2{v}}{\partial{x^2}} + \gamma vh(u) - \delta v
\end{aligned} 
\end{equation}
\label{eq:owens}
where $u$ and $v$ denote the population densities of prey and predators respectively and $f(u)$ and $h(u)$ are arbitrary positive 
functions \citep{predation}. Here $f(u)$ is a effective reproduction rate per individual. Usually, we expect $f(u)$ to be a 
decreasing function of $u$. However, sometimes, for very small $u$, the reproduction rate may decrease, say because of difficulty 
of finding a mate. If $ \frac{df(u)}{du }>0$ in some range of $u$, and an increase in density of prey leads to an increase in 
reproduction rate per individual, this is called the \emph{Allee effect}. Owen and Lewis showed that, if the dynamics is described 
by Eq.~(9), and there is no Allee effect, the velocity of prey front is not affected by the presence of predators. For the CE 
problem, increasing density of prey can not increase the reproductive rate, and there is no Allee effect. Hence, the surprise.

Figure 14 is a plot of CE and Eden fronts at values $p=0.6, 0.8$ and $1$. We can see that the centres of mass of the Eden and CE 
fronts coincide at $p=1$, and move at the same velocity. However, at $p=0.6$, the CE front is significantly slower. The inset of 
Fig.~14 further reveals that the velocity of the fronts in the pinned regime is not linear in $p$.  Anyhow, for $p\leq p_c$
the CE fronts cannot propagate any more, and thus the front speed must converge to zero for $p\to p_c$. 
What could be the reason for the this discrepancy, and the inapplicability of the Owen and Lewis analysis for CE percolation?

\begin{figure}
    \centering
    \includegraphics[scale=0.33]{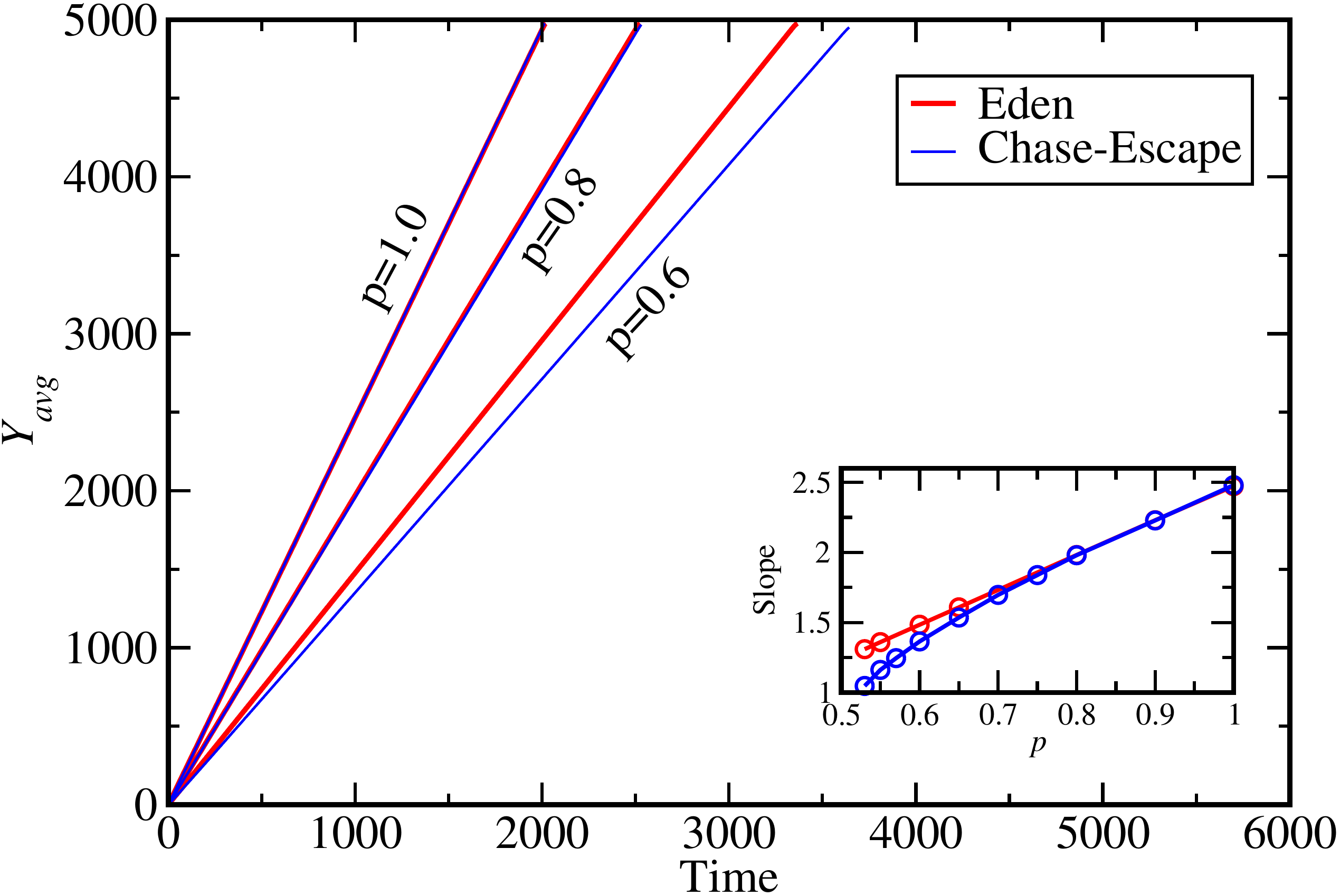}
    \caption{CE and Eden fronts at values $p=0.6, 0.8$ and $1$. We can see that the Eden and CE fronts coincide at $p=1$. As 
	pointed out in Fig.~13, the CE front is significantly slower than the Eden front at $p=0.6$.  The inset 
	is a plot of  the velocities of the CE and Eden fronts as a function of $p$. Clearly,  the velocity of the CE 
	front in the pinned regime is not linear in $p$.}
    \label{fig:my_9}
\end{figure}

\begin{figure}
    \centering
    \includegraphics[scale=0.33]{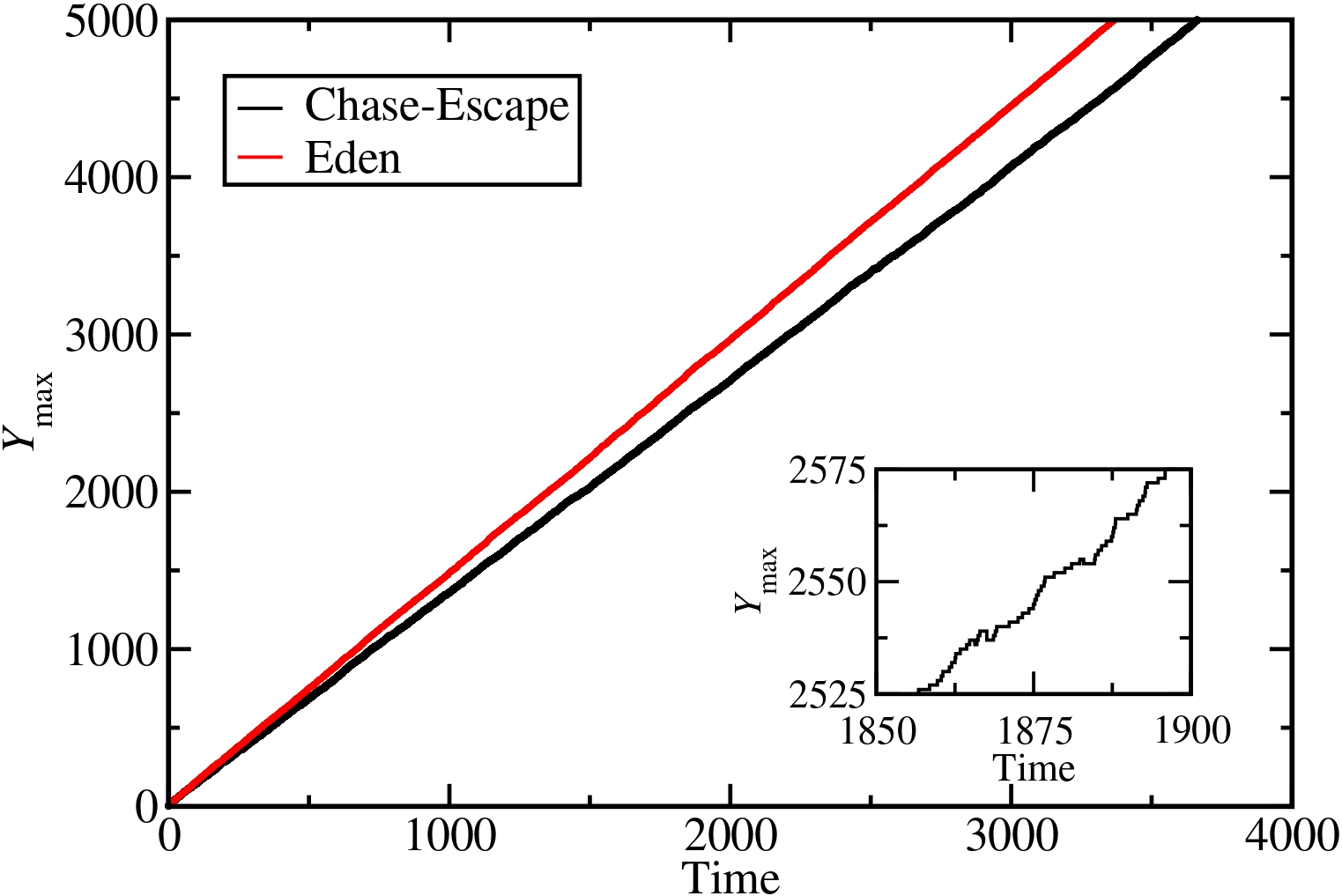}
    \caption{Two trajectories of the maximum extent of infection at $p=0.6$ for a single realization of the Eden front and the 
	red front in CE  percolation are plotted. The inset shows a zoomed-in picture of the trajectory of the CE 
	front, where  the maximum position of a red infection is seen to drop quite often.}
    \label{fig:my_10}
\end{figure}

\begin{figure}
    \centering
    \includegraphics[scale=0.33]{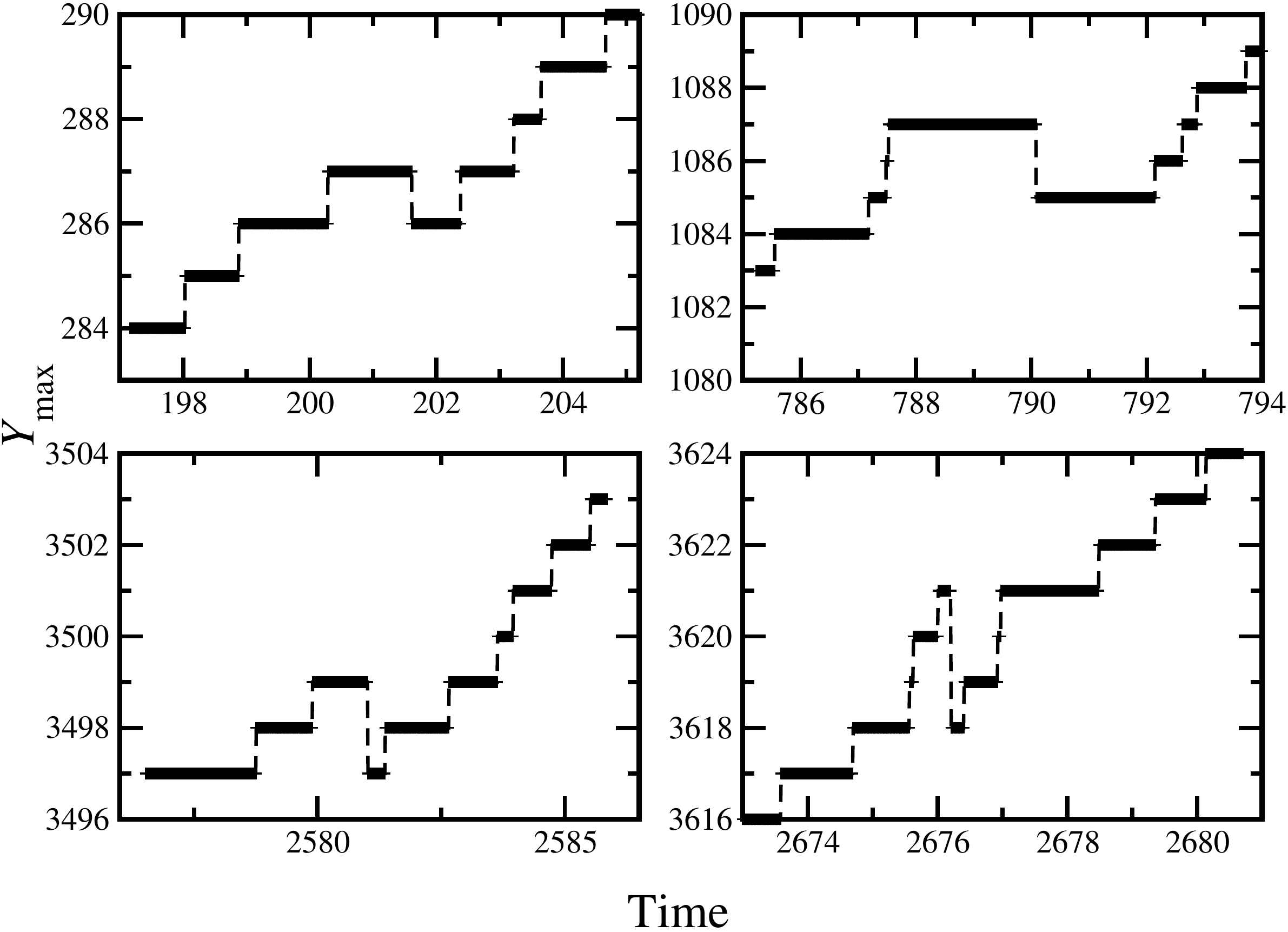}
    \caption{Four instances from the trajectory in Fig.~15 where there is a drop in the maximum position of the red front.}
    \label{fig:my_11}
\end{figure}

\begin{figure}
    \centering
    \includegraphics[scale=0.34]{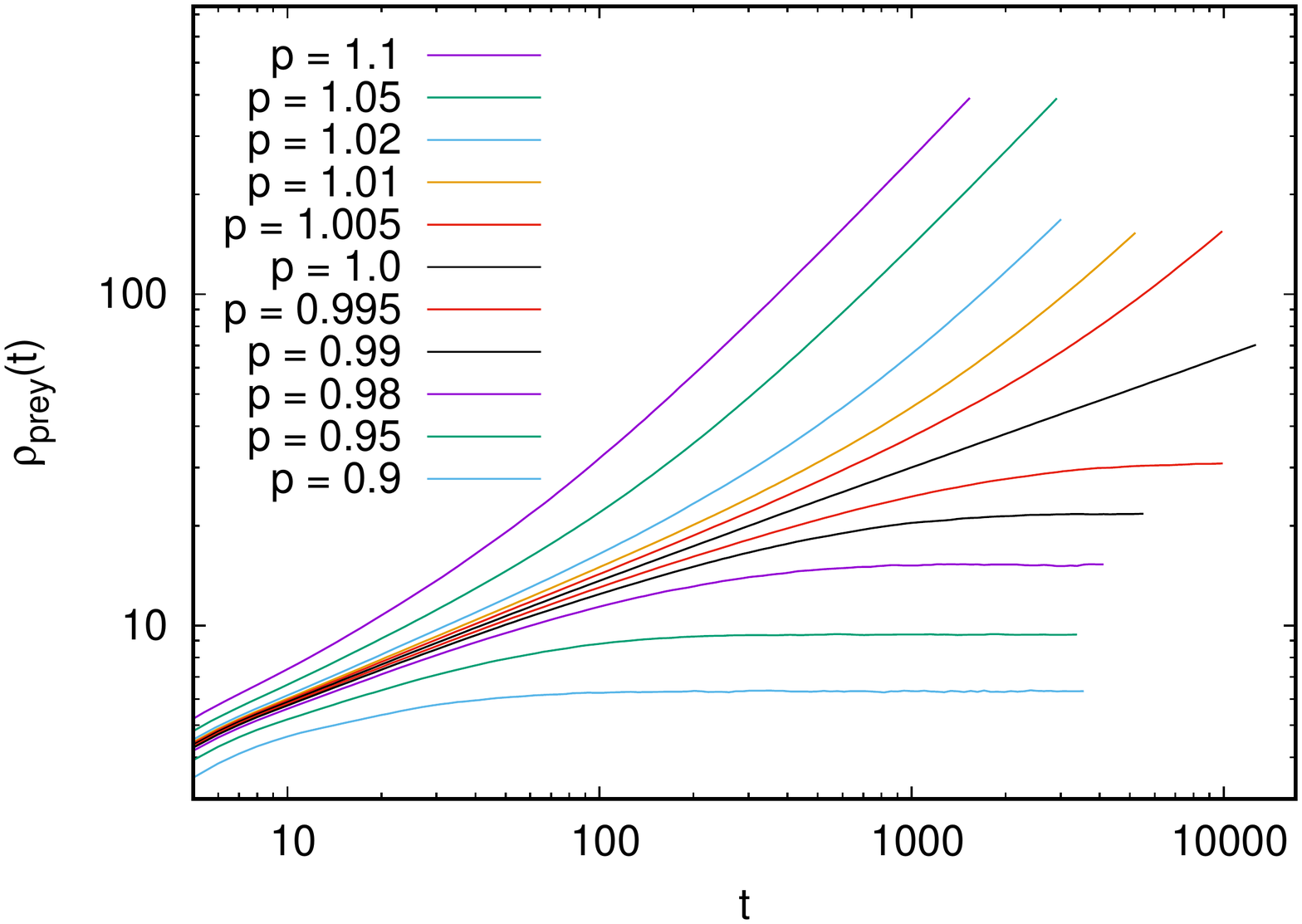}
    \caption{  Density of red sites, $\rho_{\rm red}(t)$, per unit $x$, for values of $p$ ranging from $0.9$ to $1.1$. 
	For $p_c< p <1$, $\rho_{\rm red}(t)$ tends to a constant, and for $p>1$, it grows linearly with time. At the critical 
	point $p=1$ (black curve), $\rho_{\rm red}(t) \sim t^{0.333(3)}$.}
    \label{frontwidths}
\end{figure}

\begin{figure}
    \centering
    \includegraphics[scale=0.34]{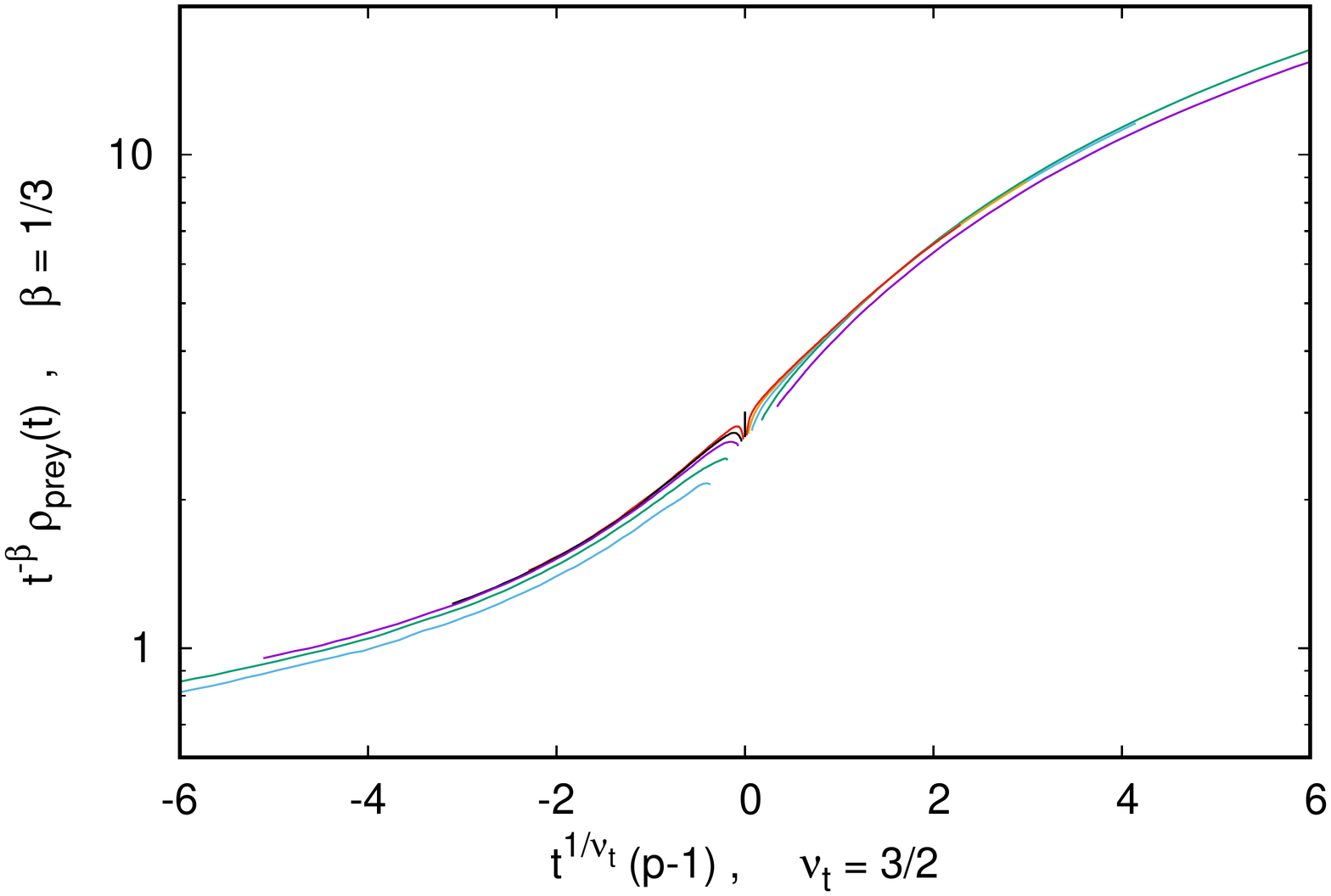}
	\caption{  Scaling collapse for  the data shown in Fig. \ref{frontwidths}. We use the scaling ansatz 
	$\rho_{\rm red}(t) = t^\beta G[(p-1)t^{\frac{1}{\nu_t}}]$ where $\beta = \frac{1}{3}$. This suggests that 
	$\nu_t = \frac{3}{2}$, and we demonstrate a good scaling collapse with these exponents. To reduce small-$t$
	corrections, only data for $t\geq 7$ are plotted.}
\end{figure}

\begin{figure*}
{\epsfig{file=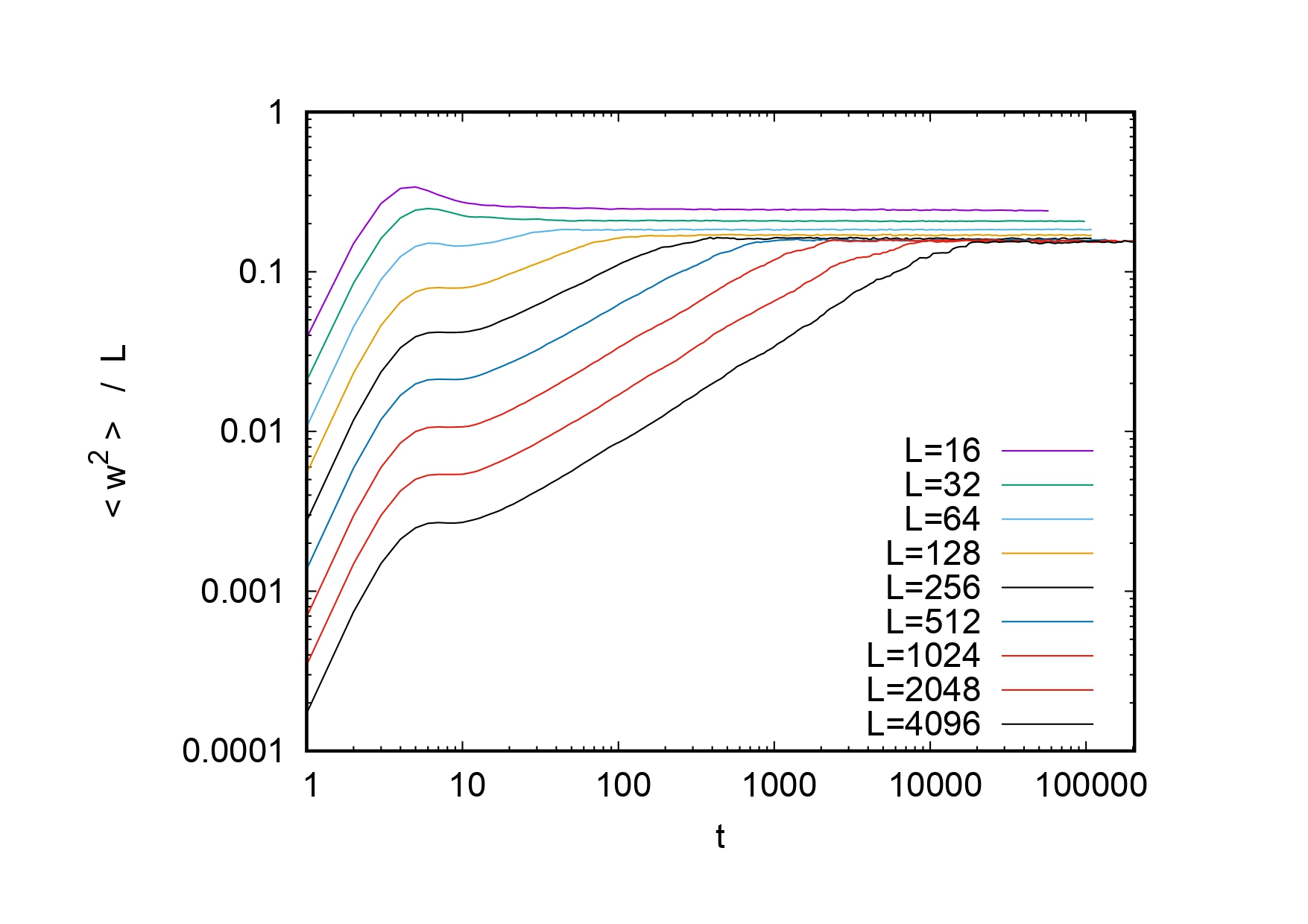,width=8.6cm}}
{\hspace*{0.5cm}\epsfig{file=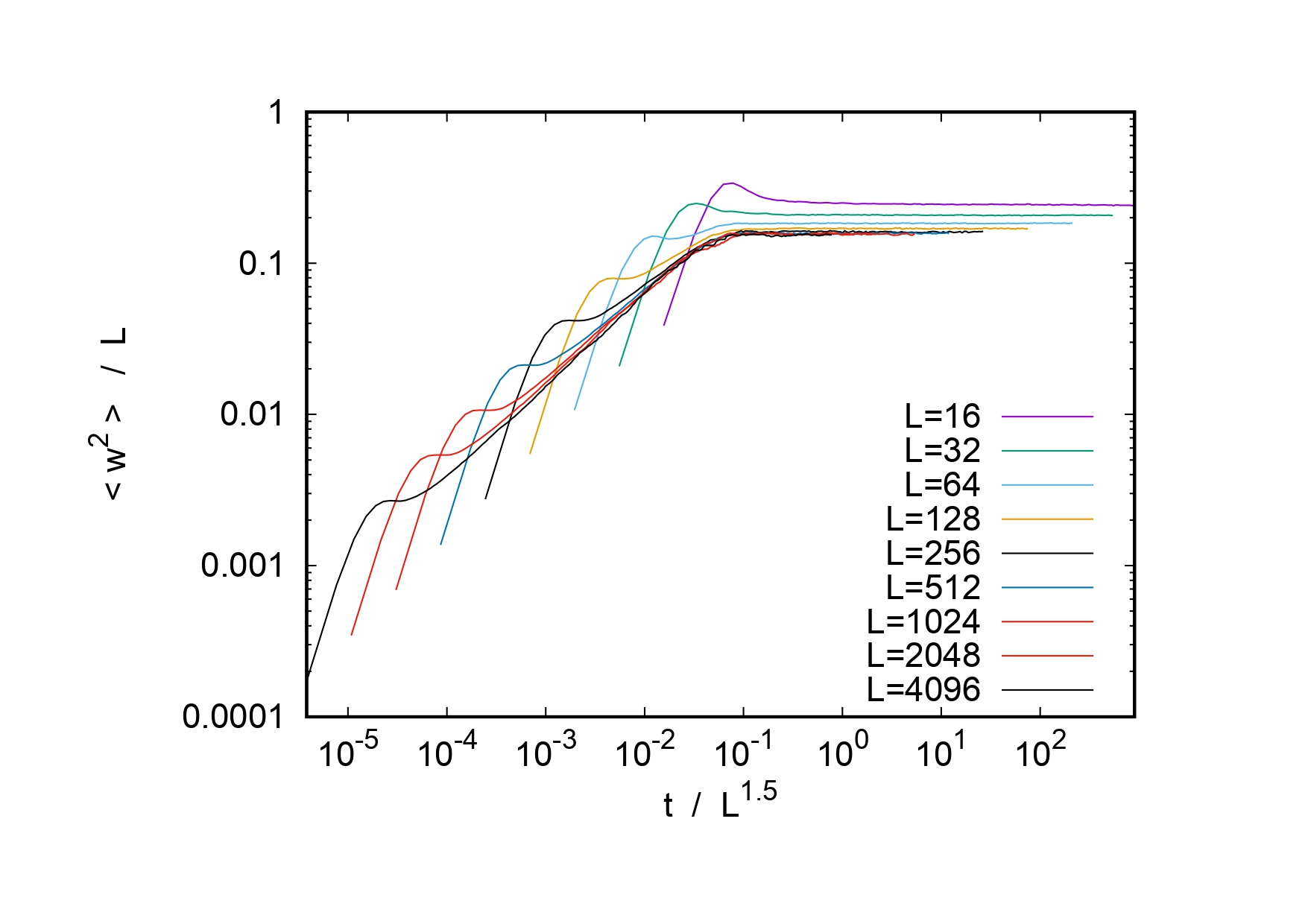,width=8.6cm}}
\caption{Scaled variance as a function of time at $p=1$. (Left) Scaling of the variance by  $L^{2\alpha}$, with $\alpha=1/2$.
	leads to a good data collapse for the asymptotic variance of the prey front. (Right) Scaling of time by $L^{z}$, with 
	dynamical exponent $z=1.5$. The exponents $\alpha$ and $z$ are consistent with the KPZ universality class.  }
\end{figure*}

A clear difference between the CE and Eden fronts is that the CE front consists of inactive parts (sites occupied by predators) 
that can only grow if there is a prey adjacent to it. In the CE front, it is also possible that the prey at a local 
maximum of the front is eaten up by a predator, and the front thereby {\em recedes}. 
This contributes to the CE front velocity becoming less than the Eden front velocity. In Fig.~15, we plot the maximum $y$ 
coordinate of the red particles ($Y_{\rm{max}}$) at $p=0.6$ for Eden front and the prey front in CE. Inset shows a zoomed 
in picture of the trajectory of the prey front where it is clearly visible that the maximum position of does drop quite often. Figure 16 shows four other instances from the trajectory of the maximum prey front where the drop is clearly visible. At $p=0.6$, the backward motion occurs roughly 1\% of the time on a lattice of size $500 \times 5000$.

We believe that the reason for this is the discrete nature of the model, which is not correctly captured in the coarse-grained continuum equations of the type described in Eq.~(9).  In the coarse-grained continuum description, $u(x)$ has a rapidly decaying tail of the distribution for large $x$, but $u(x)$ is always non-zero at finite $x$, and there is no rightmost particle. But in the discrete model there is, and there is a non-zero probability that  this  gets eaten up by the predators. These events  lead to an overall decrease in the velocity of the prey front.  Thus the discrete nature of the fluctuations of the front leads to a macroscopic observable consequence, which has also been noted in the context of the Fisher-KPP fronts \citep{slow}. The presence of predators leads to a  further decrease in  the front velocity. The continuum description fails to capture this effect. In regions $Q_2$, $Q_3$ and $Q_4$ in Fig. 6, where the predator  and prey fronts are far from each other, the Owen and Lewis analysis remains valid.

In Fig.~17, we plot the  mean density of prey $\rho_{\rm red}(t)$ per unit $x$, as a function of time, for the values 
ranging from $p=0.9$ to and $p = 1.1$. As expected, at $p  <1$, the process is in the pinned regime and 
 $\rho_{\rm red}(t)$ tends to a finite constant. At $p  >1$,  $\rho_{\rm red}(t)$ grows linearly with 
time. However, at the critical point $p=1$, by continuity arguments, one would expect that 
 $\rho_{\rm \rm red}(t) \sim t^{\beta}$ with $\beta<1$ . This expectation is verified in our numerical simulations, 
and we  find $\beta = 0.333(3)$.  To further explore the critical behavior near $p=1$, we use a scaling ansatz for 
$\rho_{\rm red}(t)$ of the form 
  \begin{equation}
	  \rho_{\rm red}(t) =  t^{\beta} G[(p-1) t^{1/\nu_t}],      \label{depinning}
   \end{equation}
with a scaling function $G(z)$ which is a finite constant at $z=0$, and has a power law scaling as $z \to \pm \infty$. 
We know that, as $t \to \infty$, $\rho_{\rm red}(t)$ tends to a  constant for $p_c<p<1$, and $\rho_{\rm red}(t) \sim (p-1)t$
for $p>1$. This would thus give us that $1/\nu_t  = 1-\beta$, implying that $\nu_t = \frac{3}{2}$. In Fig.~18, we show a 
scaling collapse for $p$ near 1, with the exponents $\beta = \frac{1}{3}$, and $\nu_t = \frac{3}{2}$.

Furthermore, for both regimes -- $p>1$ and $p<1$, one would expect that the fluctuations in the  CE fronts are captured by KPZ theory, and the process falls in the KPZ universality class \citep{kpz, kpz2}. For $p>1$ this is certainly true, because  in this regime, fluctuations of both fronts scale as  interfaces of Eden clusters.

 For $p=1$ this is not so obvious. We thus  studied the fluctuations of the CE front at the critical value $p=1$. In Fig.~19, we plot the scaled variance of the prey front as a function of time for lattice sizes $L \times 10000$, with $L$ taking values ranging from $16$ to  $4096$. The left panel is a plot of the scaling of the variance by  $L^{2\alpha}$, with $\alpha=1/2$ which leads to a good data collapse for the asymptotic variance of the prey front. For the dynamical scaling, we plot in the right panel, the scaled variance of the red front, as a function of scaled time where time has been scaled by a factor of $L^z$ with $z=1.5\pm 0.05$. The exponents thus obtained are consistent with the KPZ universality class ($\alpha =  1/2$, $z=1.5$), thus leading to the conclusion that though the depinning transition takes place at $p=1$, the critical exponents corresponding to the surface fluctuations are not affected by this transition.

\section{Summary} 
In summary, we studied the CE percolation on the $2$-$D$ square lattice. We estimated the critical probability for the survival-extinction phase transition in this model from Monte Carlo simulations, and our final estimate is that  $p_c=0.49451 \pm 0.00001$ and we can thus say with fair confidence that $p_c \neq 1/2$. From the finite size scaling analysis near this transition, we find that the correlation length exponent $\nu$ is equal to  the  $2$-$D$  undirected percolation value $4/3$ within our error bars. We also point out that the subcritical phase of CE percolation differs from the usual percolation problem, as in CE,  the cluster size distribution is not an exponentially-tailed distribution for any non-zero $p$. 

For $p$ just above $p_c$, we studied the direction dependence of the survival probabilities with the point initial condition,  and our data is fully consistent with  full rotational invariance at the critical point. Our  simulations show that in the regime $p_c < p <1$, the prey and predator fronts move with a common velocity, that is strictly less than than the velocity prey would have if predators were absent. We note that this slowing is explicitly  due to  the predators, and in addition, is due to the  decrease  in front velocity that occurs  due to fluctuations caused by the discrete nature of particles, which is also not correctly captured in the Owen-Lewis type  partial differential evolution equations without noise. We suggest that this effect depends crucially on the discrete nature of fluctuations of the advancing front. Sometimes the leading prey on the front get eaten by predators, and thus the maximum position of the prey front is a non-monotonic function of time.

 We defined a generalization of the model with discrete time, characterized by two parameters $(p_1,p_2)$. In this generalization, the CE-transition is seen along a line in the $(p_1,p_2)$ plane. One end point of the line is exactly mapped to the standard bond-percolation problem, and the other end corresponds to the original CE-model. The critical behavior is expected to be in  the standard percolation universality class along the entire line. Looking at the direction-dependent properties, we can identify other critical lines in this generalized parameter space that show  criticality corresponding the directed percolation universality class.

There are several questions that still remain unanswered about CE percolation. It would be nice to extend the results about the ladder to all graphs with a finite width, which are essentially one-dimensional, and prove that here $p_c$ is always $1$. Also, it would be nice to have a rigorous proof that for a regular $d$-dimensional lattice $p_c$ for CE percolation, $p_c$ is strictly less than 1. More generally, determination of the exact front velocity or critical probability for some non-trivial non-tree graph is an open problem.  For the generalized two-parameter model, several questions seem interesting to explore further.  One would like to understand why the phase boundary $AE$ in Fig. 6 is nearly straight. Is it everywhere convex?  The model shows rich interplay of percolation, directed percolation, pinning-depinning and active-absorbing transitions, and seems interesting for further study.

CE percolation is somewhat similar to the SIR model,  which belongs to the same universality class as undirected percolation 
\citep{siruniv}. In the SIR model, each site is either Susceptible, Infected or  ``Removed'' (i.e., dead or recovered and immune). 
Infected sites can infect neighbouring sites that are susceptible and each infected site can  die/recover independently. In some 
variants of this process, recovered sites become susceptible again with  partial immunity  (if they do not acquire any
immunity, then the resulting SIS model is in a different universality class). In CE percolation, the vacant sites can be thought 
of as susceptible, sites occupied by prey particles as infected and the ones occupied by predators as  removed. The key 
difference between the two models however is that in the SIR process, the  death/recovery of each infected particle is 
 spontaneous and  independent of others whereas in CE, the recovery can occur only at a site neighboring a recovered site, and the predators always form a single connected cluster. If one considers a variant of the CE problem, where the predators can die with a small rate $\delta$, then eventually, we have a unique absorbing state for $p$ below some $p_c^*$ (possibly different from the $p_c$ determined in this paper) and for $p>p_c^*$, we would have coexistence between the prey and predators. We would expect the survival-extinction transition at this $p_c^*$ to be in the universality class of (2+1) dimensional directed percolation, according to the Janssen-Grassberger conjecture \citep{dpc1,dpc2, hh1}. Such a crossover is also seen in the SIRS model on the square lattice, where recovered nodes can again become susceptible at a finite rate \citep{siruniv2}. We also expect a similar crossover if we allow the predators to diffuse on to empty sites at a small rate, as the extinction of the prey would stop the predator population from growing, and the existing predators would diffuse to infinity. Developing a better understanding of this crossover seems like an interesting avenue for future research.

\section*{Acknowledgements} {Aanjaneya Kumar would like to acknowledge the Prime Minister's Research Fellowship  of the government of India for financial support.}  

This paper is dedicated to the memory of Dietrich Stauffer, who was a friend, and collaborator of two of us.  Both DD 
and PG fondly remember his wit and wisdom, his dedication to physics, and his ever helpful nature. It has been a privilege 
knowing him.

\end{document}